\documentclass[12pt,preprint]{emulateapj}
\shorttitle{A debris disk study of Praesepe}
\shortauthors{G\'asp\'ar et al.}
\usepackage{lscape}

\begin{document}

\title{The low level of debris disk activity at the time of the Late Heavy Bombardment: a {\it Spitzer} study of Praesepe}

\author{A.\ G\'asp\'ar\altaffilmark{1}, G.\ H.\ Rieke\altaffilmark{1}, K.\ Y.\ L.\ Su\altaffilmark{1}, Z.\ Balog\altaffilmark{1}, 
D.\ Trilling\altaffilmark{1,2}, J.\ Muzzerole\altaffilmark{1,3}, D.\ Apai\altaffilmark{1,3}, and Kelly, B.\ C.\altaffilmark{4,5}}
\altaffiltext{1}{Steward Observatory, University of Arizona, Tucson, AZ 85721}
\altaffiltext{2}{Department of Physics and Astronomy, Northern Arizona University, Flagstaff, AZ 86011}
\altaffiltext{3}{Space Telescope Science Division of ESA, STScI, 3700 San Martin Drive, Baltimore, MD 21218}
\altaffiltext{4}{Harvard-Smithsonian Center for Astrophysics, 60 Garden St, Cambridge, MA 02138}
\altaffiltext{5}{Hubble Fellow}
\email{agaspar@as.arizona.edu}

\begin{abstract}
We present 24 $\micron$ photometry of the intermediate-age open cluster Praesepe.
We assemble a catalog of 193 probable cluster members that are detected in optical databases, the Two Micron All Sky Survey (2MASS),
and at 24 $\micron$, within an area of $\sim 2.47$ square degrees. Mid-IR excesses indicating debris disks are found for one early-type
and for three solar-type stars. Corrections for sampling statistics yield a 24 $\micron$ excess fraction (debris disk fraction) of $6.5\pm4.1$\%
for luminous and $1.9\pm1.2$\% for solar-type stars. The incidence of excesses is in agreement with the decay trend 
of debris disks as a function of age observed for other cluster and field stars. The values also agree with those for older stars, 
indicating that debris generation in the zones that emit at 24 $\micron$ falls
to the older 1-10 Gyr field star sample value by roughly 750 Myr.

We discuss our results in the context of previous observations of excess fractions for early- and solar-type stars. 
We show that solar-type stars lose their debris disk 24 $\micron$ excesses on a shorter timescale than
early-type stars. Simplistic Monte Carlo models suggest that, during the first Gyr of their evolution, up to 
15-30\% of solar-type stars might undergo an orbital realignment of giant planets such as the one thought 
to have led to the Late Heavy Bombardment, if the length of the bombardment episode is similar to the one
thought to have happened in our Solar System.

In the Appendix, we determine the cluster's parameters via boostrap Monte Carlo isochrone fitting, yielding an age of
757 Myr ($\pm$ 36 Myr at 1$\sigma$ confidence) and a distance of 179 pc ($\pm$ 2 pc at 1$\sigma$ confidence), not allowing
for systematic errors.

\end{abstract}


\section{Introduction}

Stars generally form with an accompanying circumstellar disk. 
Planets can grow from this primordial disk over a few to a few tens of
Myr. The {\it Infrared Astronomy Satellite} ({\it IRAS}) detected infrared excess emission from 
disks around stars with ages much older than the 
clearing timescales of protoplanetary circumstellar disks \citep{aumann84}. These excesses arise 
from second-generation "debris disks" that are the results of collisional cascades 
initiated by impacts between planetesimals and of cometary activity \citep{backman93}. 
The micron-sized dust grains in debris disks are heated by the central star(s) and reradiate the received energy at 
mid-infrared wavelengths. Studying this infrared emission 
lets us probe the frequency of formation of planetary systems and to track their evolution.
For example, some of the relatively prominent disks may be analogs to that in the Solar System at the 
epoch of Late Heavy Bombardment \citep[LHB; e.g.][]{gomes05,strom05}.

{\it IRAS} and {\it Infrared Space Observatory}
({\it ISO}) observations of debris disks suggest that the excess rate steadily declines
with stellar age, indicative of stars losing these disks within a few hundred million
years \citep{habing01,spangler01}. A theoretical model that involved delayed stirring was developed 
by \cite{dominik03} to explain this phenomenon; however, a uniform evolutionary model could not be derived. 
There were a number of reasons. The sensitivity of these instruments was often inadequate for observations down to the photospheric levels. The large 
beam sizes also occasionally confused the excesses
with background objects and/or the galactic cirrus. The Multiband Imaging Photometer for {\it Spitzer} 
\citep[MIPS;][]{rieke04} on the {\it Spitzer Space Telescope} has improved sensitivity and 
resolution in the mid-infrared and with it astronomers have been able to carry out more detailed statistical studies of debris 
disks at a wide range of stellar ages and spectral types.

\cite{rieke05} observed a large sample of nearby A-type field stars with {\it Spitzer}, which combined with
existing {\it IRAS} and {\it ISO} data definitively demonstrated that the frequency of debris disk excesses declines
with age and that the disk properties vary at all ages. Even by probing excesses down to 25\% above the photospheric level, \cite{rieke05}
found that some stars at ages of only 10-20 Myr do not show any signs of excess. These results were confirmed by \cite{su06}. 
This behavior implies a very fast clearing mechanism for disks around some of these stars, or perhaps that they form with only very low mass disks. 
The models of \cite{wyatt07} provided a first-order explanation in terms of a steady state evolution of the debris disks from a 
broad distribution of initial masses.

An important question for habitable planet search/evolution is whether the same processes occur 
for FGK-type stars. A number of surveys of solar-type stars have been conducted with {\it Spitzer}. The MIPS Guaranteed Time Observers (GTO)
team has searched $\sim 200$ field stars for excesses \citep{trilling08}, plus many hundreds of open cluster members
\citep[e.g.,][]{gorlova06,gorlova07,siegler07}. The legacy survey
by the Formation and Evolution of Planetary Systems (FEPS) group has examined 328 stars 
(both field and open cluster members)\citep{mamajek04,meyer04,meyer08,stauffer05,kim05,silverstone06}. 

\cite{trilling08} showed that solar-type stars
of age older than 1 Gyr have excess emission at 70 $\micron$ $\sim 16$\% of the time. Excesses at this wavelength are expected to arise from
Kuiper-Belt-like planetesimal regions, but with masses 10-100 times greater. \cite{meyer08}
find that 8.5-19\% of solar-type stars at ages $<300$ Myr have debris disks detectable at 24 $\micron$ and that this number gradually goes down to 
$<4$\% at older ages, augmenting work by \cite{gorlova06}, \cite{siegler07} and \cite{trilling08}. Excesses at this wavelength around solar-type
stars probe the 1-40 AU range, the asteroidal and planetary region in the Solar System.

The ideal laboratories to determine the stellar disk fractions with good number statistics 
are open clusters and associations.  To investigate the fraction of solar-type excess 
stars, the observations have to be able to detect the photospheres of the non-excess stars. The range
of distances to suitable clusters compromises the uniformity of the results. 
The survey of h and $\chi$ Persei 
\citep{currie08} could only determine the early-type star excess fraction, while that of NGC 2547 
\citep{young04,gorlova07} could only detect photospheres down to early G due to similar limits. The observations in 
M47 \citep{gorlova04} also yielded values to early G spectral type stars. The investigations of IC 2391 \citep{siegler07} 
and the Pleiades \citep{gorlova06} gave insights on debris disk evolution down as far as K spectral-type stars.

To study further the fraction of debris disks around solar-mass stars, we have observed the nearby Praesepe (M44, NGC 2632, Beehive) open cluster. 
Our observations, along with those of \cite{cieza08} on the Hyades cluster, fill the gap in previous work on debris disk fractions in the
age range of 600-800 Myr. This range is of interest because it coincides with the LHB in the Solar System.
The close proximity of the cluster ($\sim$ 180 pc) and its large number of members
ensured that good statistics would be achieved. Praesepe has been extensively studied by many groups 
\citep{kw,jones83,jones91,hambly95,wang95,kraus07}, providing a nearly full membership list to our completeness limit
of $[24] \sim 9$ mag (the brightness of a G4 V spectral-type star at the distance of the cluster). The member stars
have high proper motions ($\sim 39$ mas yr$^{-1}$), clearly distinguishing them from field stars.

\section{Observations, data reduction, and photometry}

\begin{figure}[t]
\begin{center}
\includegraphics[angle=0,scale=0.8]{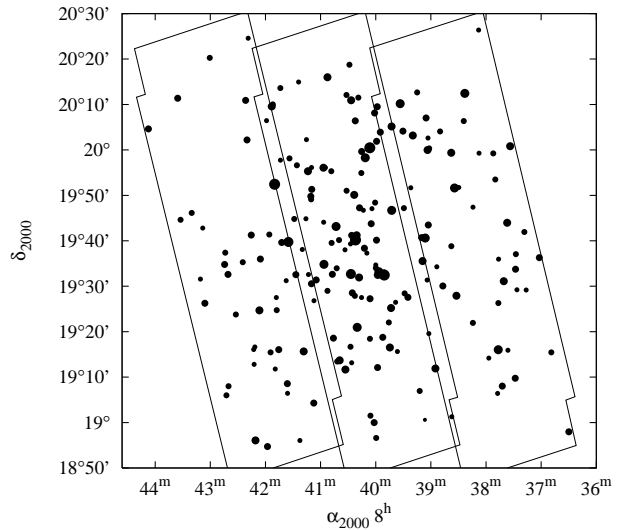}
\caption{The observed field, showing the areas covered by three scanmaps and the observed cluster member stars with sizes proportional to their brightness in [24].}
\label{fig:starcoord}
\end{center}
\end{figure}

We used MIPS to observe Praesepe as part of the GTO program PID 30429 (2007 May 30). The center part of the cluster (8$^{\rm h}$40$^{\rm m}$21$^{\rm s}$, 
19$^{\circ}$$38'$$40''$) was imaged using three scan maps (with 12 legs in a single scan map overlapping 
with half-array cross-scan). The map covers a field of $\sim 2.47 ~{\rm deg}^2$, as shown in Figure \ref{fig:starcoord}. We used medium scan mode, resulting in 
a total effective exposure time per pixel of 80 s (at 24 $\micron$). All data were processed using the MIPS instrument team Data Analysis Tool \citep[DAT,][]{gordon05} 
as described by \cite{engelbracht07}. 

Although MIPS in scan-mode provides simultaneous data from all three detectors (at 24, 70 and 160 $\micron$),
we base our study on the 24 $\micron$ channel data only. The 70 and 160 $\micron$ detectors are insensitive
to stellar photospheric emissions at the distance of Praesepe. In retrospect, the rarity of excesses in our survey is consistent with the
lack of detections at the longer wavelengths.

The initial coordinate list for the 24 $\micron$ photometry was assembled with the {\tt daofind} task under IRAF\footnote{IRAF is distributed by the National 
Optical Astronomy Observatories, which is operated by the Association of the Universities for Research in Astronomy, Inc.\ (AURA) under cooperative agreement 
with the National Science Foundation.}. We later expanded this list by visually examining the images and manually adding all sources to the list that were missed by
{\tt daofind}. Our final list for photometry contained 1457 sources. To achieve high accuracy, we performed point-spread function (PSF)--fitting photometry.
The calibration star HD 173398 was adopted as a PSF standard, with the final PSF constructed from 72 individual observations, kindly provided 
to us by C.\ Engelbracht. The standard IRAF tasks {\tt phot} and {\tt allstar} of the {\tt daophot} package were used.

\begin{figure*}[t]
\begin{center}
\includegraphics[angle=0,scale=0.6]{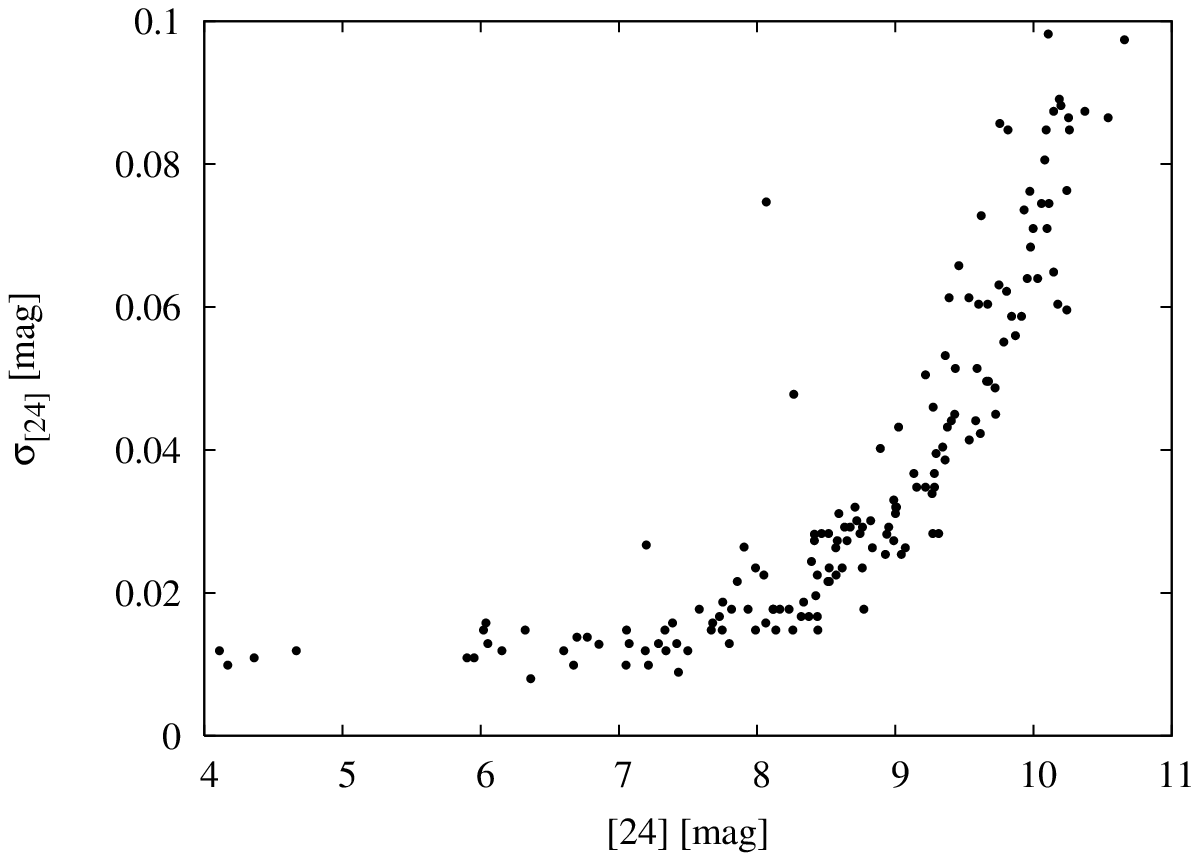}
\includegraphics[angle=0,scale=0.6]{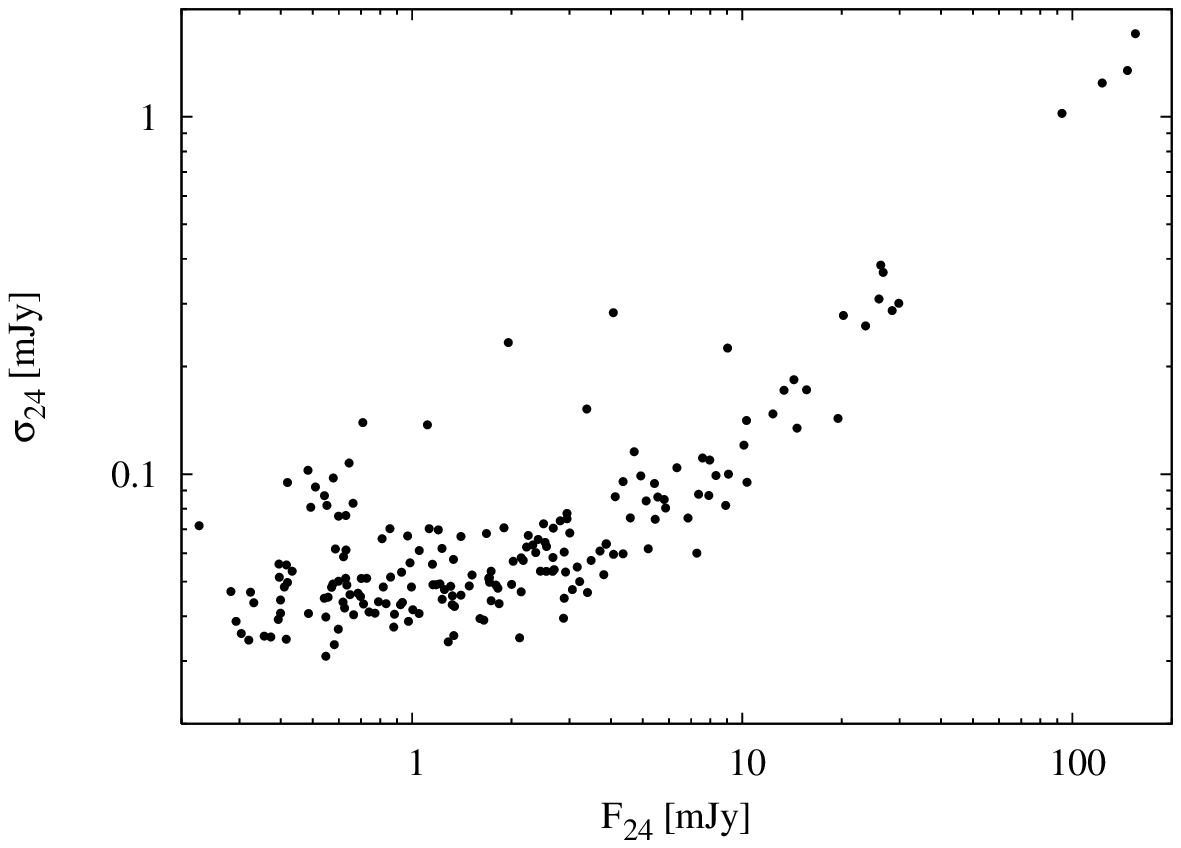}
\caption{The error of the 24 $\micron$ photometry is plotted as a function of brightness for cluster member sources. The {\it left panel} shows the flux and its 
error on the magnitude scale, while the {\it right panel} shows them in mJy flux values. All points have less than 0.1 magnitude error and nearly all stars 
brighter than $9^{\rm th}$ magnitude have errors less than $0.04$ magnitude.}
\label{fig:magerr}
\end{center}
\end{figure*}

The observed field is free of nebulosity and stellar crowding, so we were able to use a large
PSF radius of $112''$, with fitting radius of $5.7''$. The large PSF radius ensured us that the aperture correction was negligible.
The instrumental number counts were converted to flux densities with the conversion $1.068\times10^{-3} ~{\rm mJy}~{\rm arcsec}^{-2}~{\rm MIPS\_UNIT}^{-1}$ 
\citep{engelbracht07}. We then translated these values to 24 $\micron$ magnitudes 
taking 7.17 Jy  for the [24] magnitude zero point, which has error of $\pm$ 0.11 Jy \citep{rieke08}. We show the photometric error vs.\ brightness plots 
of our measurements in Figure \ref{fig:magerr}. Almost all sources brighter than 9$^{\rm th}$ magnitude ($\sim$ 1.8 mJy) have errors less than 0.04 mag ($\sim$ 0.07 mJy) and
all sources remain below errors of 0.1 mag; the average error is $\sim 5$\%. As a check, we performed independent PSF photometry with {\tt StarFinder} under 
IDL, obtaining photometry values within the errors of our IRAF photometry and with errors similar to the ones given by {\tt daophot}.

\section{Catalog surveys and the final sample}

We compiled a complete catalog for all sources in our field of view, including their optical, near infrared, and 24 $\micron$ data. We expanded this catalog
with all known cluster members outside of our field of view (naturally without [24] data). This enabled us to plot a full cluster optical color-magnitude diagram (CMD), which 
we used to confirm the cluster's age and distance (see Appendix).

\begin{figure*}[]
\begin{center}
\includegraphics[angle=0,scale=0.87]{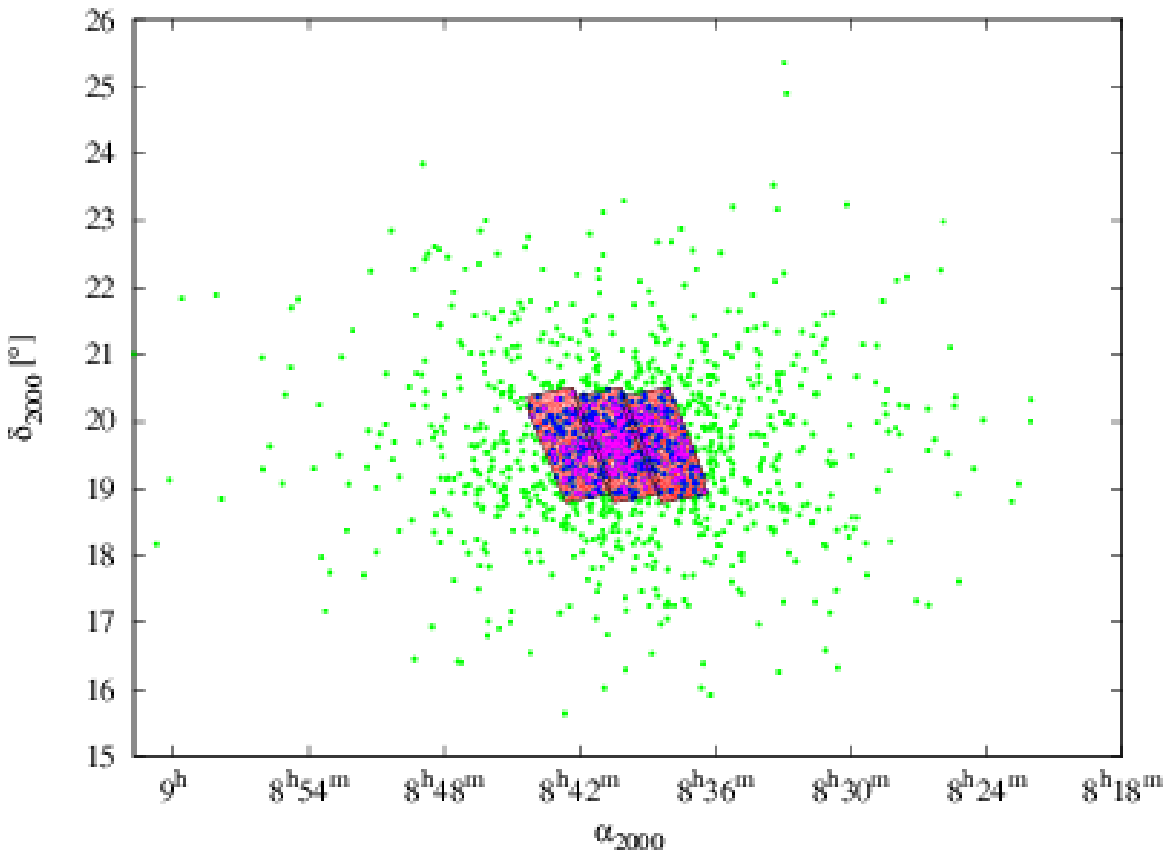}
\includegraphics[angle=0,scale=0.87]{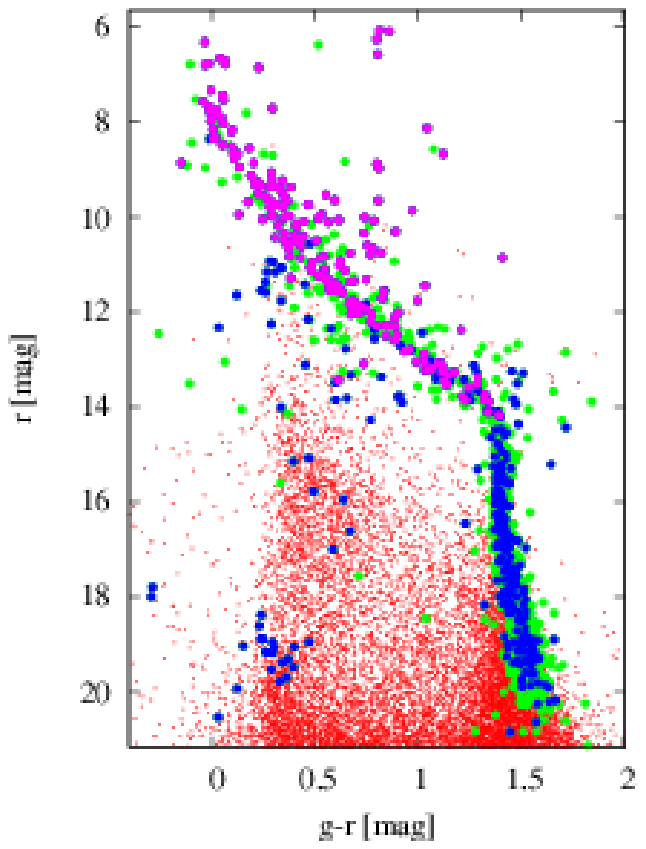}
\caption{The spatial and CMD position of the selected sources. {\it Red dots:} The combined list for all sources in our observed field; {\it Green
dots:} All cluster members outside our observed field; {\it Blue dots:} All cluster members in our observed field that could not be identified
in our 24 $\micron$ survey; {\it Magenta dots:} All cluster members that were identified in our 24 $\micron$ survey.}
\label{fig:selection}
\end{center}
\end{figure*}

Optical data for the sources were obtained from the 5$^{\rm th}$ data release of the Sloan Digital Sky Survey (SDSS),
while 2MASS provided J, H, and K$_{S}$ magnitudes. The SDSS photometry is generally unreliable for bright sources, the ones mostly detected in our MIPS survey.
To ensure we had good photometry for these sources, we collected BV data (for 356 stars altogether) using the Webda 
database\footnote{http://www.univie.ac.at/webda/}, providing an ensemble of data for 
high probability cluster members from various papers \citep{johnson52,twarog82,dickens68,lutz77,upgren79,castelaz91,mermilliod90,weis81,stauffer82,
andruk95,mendoza67,oja85}. The data downloaded from the Webda database cover the brightest magnitude range of the cluster, including stars avoided
by modern CCD observations or where they are saturated. We converted the BV magnitudes to SDSS r and g values by averaging the conversion slopes of 
\cite{jester05,jordi06,zhao06} and \cite{fukugita} and obtained
\begin{eqnarray}
g&=&(0.607\pm0.016)(B-V)\nonumber \\
&& -(0.1153\pm0.0095)+V \\
r&=&(-0.453\pm0.028)(B-V)\nonumber \\
&& +(0.1006\pm0.0131)+V.
\end{eqnarray} 
Where our calculated r or g brightnesses for the Webda catalog members differed from the SDSS data by more than 0.5 magnitude, we replaced the SDSS data with the 
calculated one. 

Cluster membership was determined by compiling all accessible databases. The largest membership lists are those of \cite{wang95} and \cite{kraus07}, which were supplemented by 
our Webda catalog search results. \cite{wang95} give a list of 924 stars, out of which we chose only 198 that are high probability members of the cluster according to 
the proper motion data in the paper. The list of \cite{kraus07} is much more robust with 1130 stars, all of which have membership probability $>50$\%; 1010 of them have 
$>80$\% membership probability. The databases (SDSS, 2MASS, Webda, \cite{wang95}, \cite{kraus07}) were cross-correlated with a maximum matching radius of $3.6''$. 
The closest member within this radius is matched as a pair and all others are added to the catalog
as new sources. The program excluded pairing members from the same catalog. Our final cluster member list contains 1281 candidates, of which 493 were in our observed field. 

After plotting the color-magnitude diagram and doing an initial isochrone fit on cluster members, we tested for bad photometry. We generated a list of all the member stars 
that were further from the isochrone sequence than 0.3 magnitude, examined all these stars for anomolies on SDSS images, and searched for BV magnitudes in Simbad. If the star 
was saturated or a calculated r, g magnitude differed from the SDSS r, g value by 0.5 magnitude or more (the same criteria as used before), we used the 
calculated value. 

\begin{figure}[t]
\begin{center}
\includegraphics[angle=0,scale=0.6]{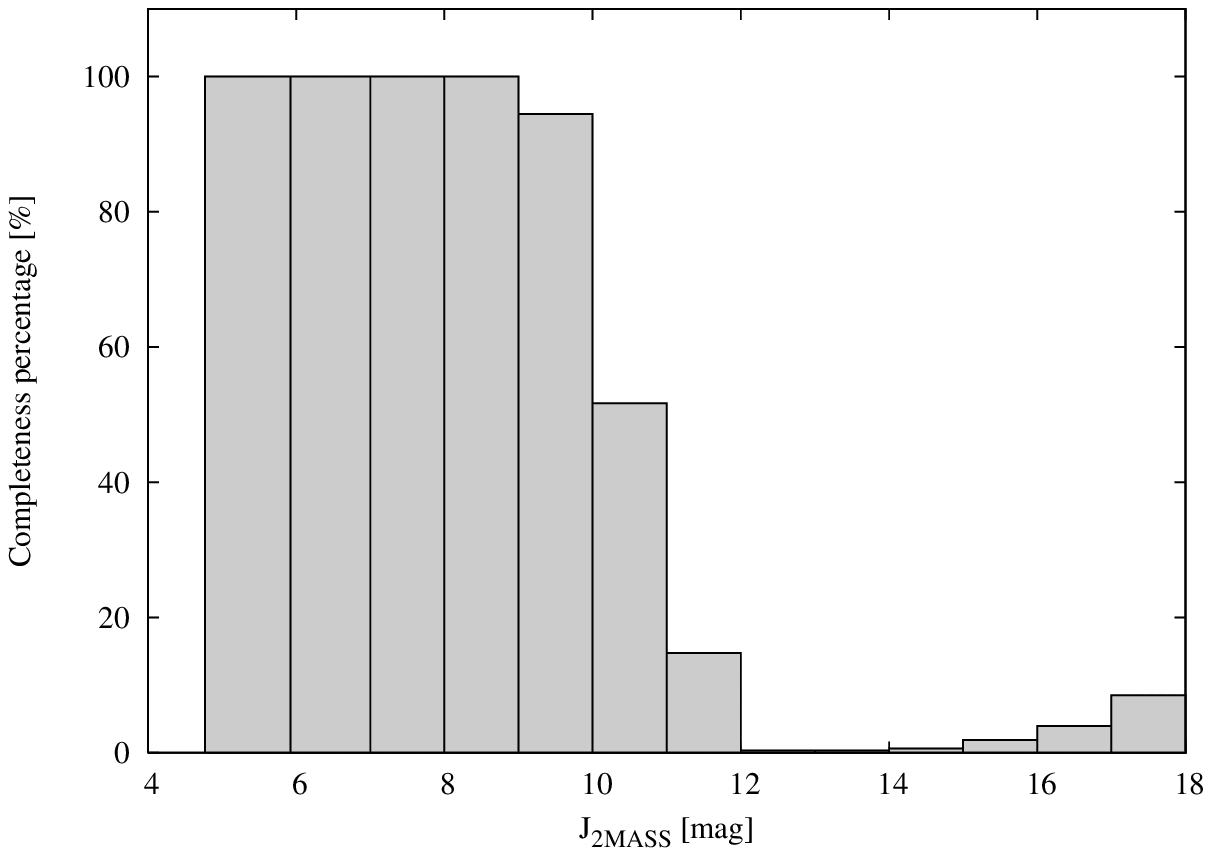}
\includegraphics[angle=0,scale=0.6]{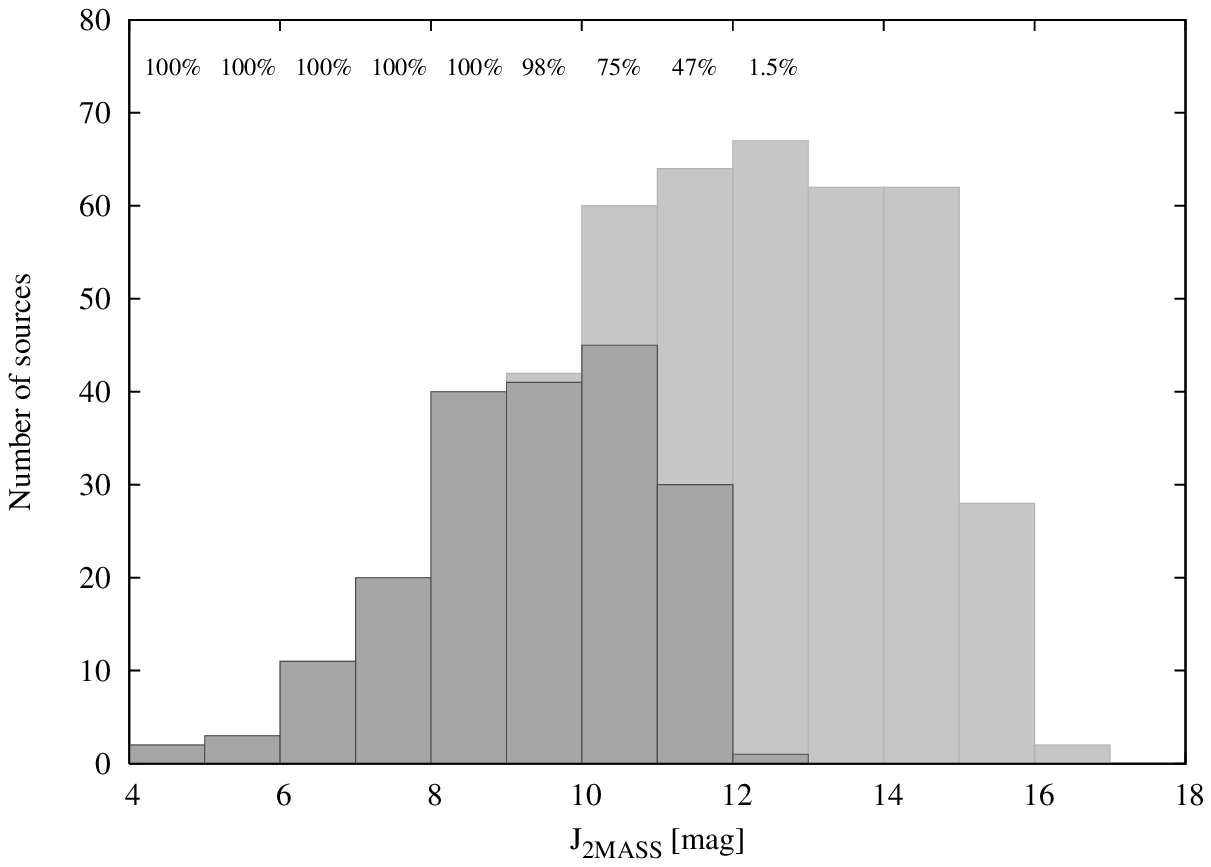}
\caption{{\it Left Panel:} The completeness limit of our survey is shown as a function of 2MASS J magnitude. We detect almost all sources brighter than 10$^{\rm th}$ magnitude,
corresponding to a $\sim$ G4 V star. For sources fainter than 14$^{\rm th}$ magnitude random associations begin to occur.
{\it Right Panel:} The total number of cluster members within our field of view (light gray) and the number of members detected (dark gray) at [24] are shown as a function of J
magnitude.}
\label{fig:completion}
\end{center}
\end{figure}

In Figure \ref{fig:selection}, we show how the selection criteria narrow the CMD, and where sources with different selection characteristics are distributed in the field.
From the 1457 sources identified in our 24 $\micron$ survey, 201 were cataloged as cluster members by previous work. Of these, 193 also have data in the optical and near 
infrared. Our survey's completeness limit compared to 2MASS is at ${\rm J}=10 ~{\rm mag}$ ($[24]\sim 9~{\rm mag}$), as is shown in 
Figure \ref{fig:completion}. This limit corresponds to a G4 V star at the distance of Praesepe. The completeness limit for the cluster member sources 
is also shown in Figure \ref{fig:completion}. Between 10$^{\rm th}$ and 11$^{\rm th}$ magnitude in J we achieve 75\% completeness for cluster members.

\begin{figure*}[t]
\begin{center}
\includegraphics[angle=0,scale=1.4]{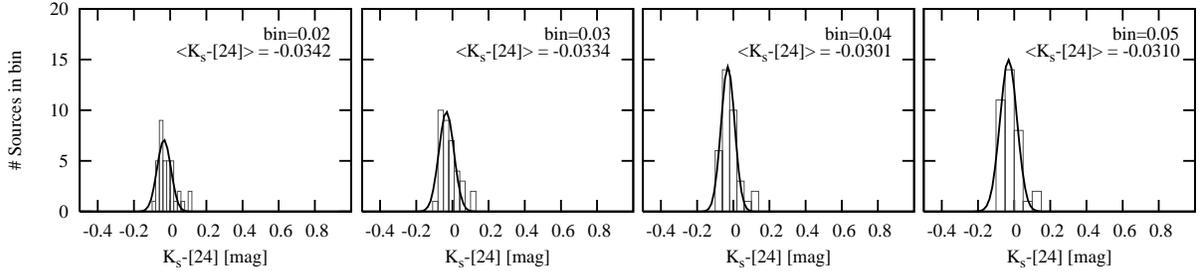}
\caption{The panels show the Gaussian fits to the number distribution of A type stars, within certain K$_S$-[24] bins.}
\label{fig:fitting}
\end{center}
\end{figure*}

For our [24] magnitude values to be comparable to the 2MASS K$_{S}$ photometry, we fitted a Gaussian to the binned number distribution of 
the K$_S$-[24] values of all member sources with r-K$_{S} < 0.8$ ($\sim$ A stars). We derived a general correction factor of -0.032$\pm$0.002 magnitude ($\sim$ 3\%) for the
[24] values. The Gaussian fits are shown in Figure \ref{fig:fitting}. This same method has been used by \cite{rieke08} to obtain the average ratio of K$_S$ to 24 $\micron$ flux 
densities. By optimizing the fit of our [24] data to the 2MASS K$_{S}$ data, we eliminated any absolute calibration offsets. The average variance of the fitted
gaussians is $\sigma=0.047$ mag, consistent with our average [24] error value of $\sim 0.05$ mag.

We summarize our [24] photometry results for the 193 cluster members that were identified in all wavelength regions in Table \ref{tab:solutions}. The first column of the table
gives our designated number, while the coordinates are that of the 24 $\micron$ flux source. As a source/coordinate comparison we also list the 2MASS source associated with 
the 24 $\micron$ emission. The table contains the K$_{S}$ adjusted [24] magnitude, the original flux values (in mJy) and the "best" r and g photometry value. Cluster membership
probability is shown by either the proper motion of the source or by the \cite{wang95} catalog number of the source. Sources that are missing both values were listed as
cluster members either in the Webda database or in \cite{kraus07}.

\section{Results}

In this section we present our results on the debris disk fraction we observed in Praesepe and place it in context with previous results on the evolution of
debris disks. There are two basic methods to detect 24 $\micron$ excess. The first is to use a color-color diagram, with one of the colors determining
the stars' spectral type and the other being K$_{S}$-[24]. The r-K$_{S}$ color is ideal to differentiate spectral types, while the 
K$_{S}$-[24] color depends only weakly on the spectral type of the star since both wavelengths fall on the Rayleigh-Jeans part of the spectral energy 
distribution (SED) for all sources hotter than early M type (T$_{\rm eff} > 3200$ K) \citep{gautier07}. For non-excess stars the K$_{S}$-[24] color should stay close to zero. 
Any excess measured in K$_{S}$-[24] is most likely caused by circumstellar material.
 
The second method is to fit the observed optical and near-infrared photometry with theoretical SEDs based on stellar photosphere models. Excesses are revealed if the 
24 $\micron$ flux density is significantly greater than the predicted flux.

\subsection{Color-color selection}

\begin{figure}[t]
\begin{center}
\includegraphics[angle=0,scale=1.3]{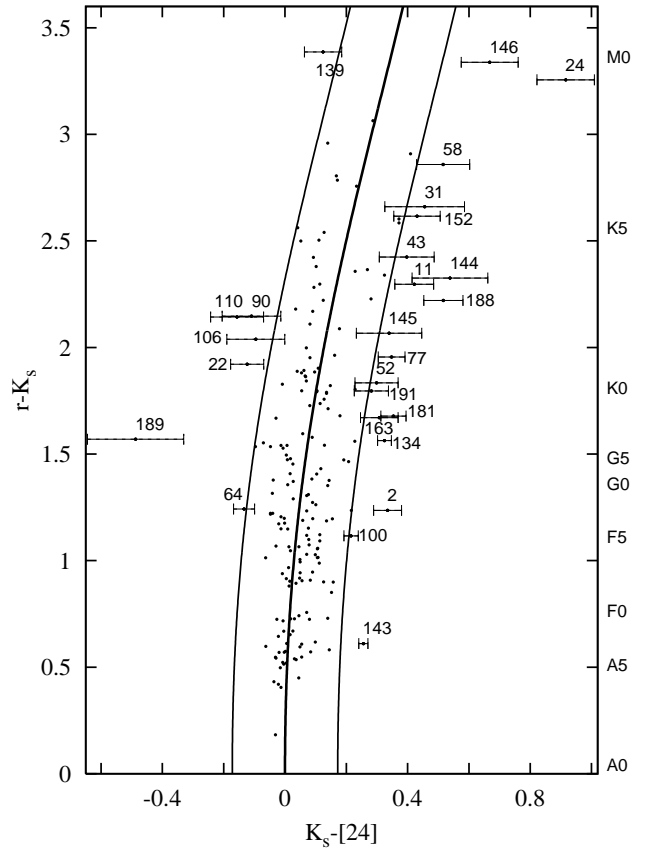}
\caption{The color-color plot for the cluster members with photometric measurements in r, K$_{S}$, and [24]. The 1$\sigma$ measurement error in [24] is plotted for stars
that are outside of the trend curve. The nomenclature is from Table \ref{tab:solutions}.}
\label{fig:2color}
\end{center}
\end{figure}

We used the color-color diagram shown in Figure \ref{fig:2color} as our primary method to identify sources as excess candidates. We plot all cluster 
members that have magnitude values in r, K$_{S}$, and [24], 193 sources altogether. 

\cite{gautier07} show the trend of K$_{S}$-[24] photospheric color with spectral type for stars of low effective temperature.
The empirical locus of stars on the color-color plot in Figure \ref{fig:2color} was derived by fitting a curve to a sample of field stars (from \cite{gautier07} and \cite{trilling08}).
We then converted the fitted V-K$_{S}$ colors to r-K$_{S}$ colors through conversion tables in
\cite{allen} and \cite{kraus07}. Our final color-color curve for r-K$_{S}$ vs.\ K$_{S}$-[24] for main-sequence (MS) stars is:
\begin{eqnarray} 
{\rm K}_{S}-{\rm [24]} &=& 3.01\times10^{-5}({\rm r}-{\rm K}_{S})+0.0233({\rm r}-{\rm K}_{S})^2 \nonumber \\
&&+0.0072({\rm r}-{\rm K}_{S})^3-0.0015({\rm r}-{\rm K}_{S})^4.
\end{eqnarray}
In Figure \ref{fig:2color}, we plot this curve and the 3$\sigma$ average confidence level for our photometry in [24] ($\sim 0.15$ mag) (the errors of the curve itself 
are minor compared to the photometric errors). The 
majority of the stars ($> 86$\%) lie within this band. The errors plotted for the stars outside of our MS fitted curve are the 1$\sigma$ errors
in our [24] photometry. To use the K$_{S}$-[24] color as an excess diagnostic tool, one must make sure that the K$_{S}$ magnitude
is truly photospheric. We examined the J, H, and K$_{S}$ fits to theoretical SEDs \citep{kurucz11} and concluded that all K$_{S}$
magnitudes are truly photospheric; the largest difference (from our debris disk candidate sample introduced later) is in the case of star 143, where the 
measured value is above the predicted SED value by 5.6\%.

\setcounter{table}{1}
\begin{deluxetable*}{lrrrrrrrrr}
\tablewidth{0pt}
\tablecolumns{10}
\tablecaption{The probabilities of chance alignments for our sources with background galaxies as a function of [24] brightness.\label{tab:chance}}
\tablehead{
\colhead{[24] bin} & \colhead{N$_{\ast}$} & \colhead{Flux} & \colhead{Excess} & \colhead{N$_{\rm galaxies}$} & \multicolumn{5}{c}{$P$ of at least $n$ chance alignments} \\
\colhead{[mag]} & \colhead{[\#]} & \colhead{[mJy]} & \colhead{[mJy]} & \colhead{[sr$^{-1}$]} & \colhead{[0]\tablenotemark{$\ast$}} &
\colhead{[1]\tablenotemark{$\dagger$}} 
& \colhead{[2]\tablenotemark{$\dagger$}} & \colhead{[3]\tablenotemark{$\dagger$}} & \colhead{[4]\tablenotemark{$\dagger$}}}
\startdata
4-5   & 4  & 125.900 & 16.245 & 2$\times10^{4}$ & 99.99\% &  0.01\% & $\sim$0\% & $\sim$0\% & $\sim$0\% \\
5-6   & 2  &  50.122 &  6.467 & 7$\times10^{4}$ & 99.98\% &  0.02\% & $\sim$0\% &         - &         - \\
6-7   & 11 &  19.954 &  2.575 & 4$\times10^{5}$ & 99.54\% &  0.46\% & $\sim$0\% & $\sim$0\% & $\sim$0\% \\
7-8   & 26 &   7.944 &  1.025 & 1$\times10^{6}$ & 97.46\% &  2.54\% &    0.04\% & $\sim$0\% & $\sim$0\% \\
8-9   & 48 &   3.162 &  0.408 & 7$\times10^{6}$ & 72.35\% & 27.65\% &    3.80\% &    0.01\% & $\sim$0\% \\
9-10  & 53 &   1.259 &  0.162 & 4$\times10^{7}$ & 12.60\% & 87.40\% &   60.26\% &   31.93\% &   13.14\% \\
10-11 & 48 &   0.501 &  0.065 & 8$\times10^{7}$ &  2.47\% & 97.53\% &   87.34\% &   69.30\% &   47.06\% \\
11-12 & 1  &   0.199 &  0.026 & 1$\times10^{8}$ & 91.20\% &  8.80\% &         - &         - &         -
\enddata
\tablenotetext{$\ast$}{The probability that none of the cluster member sources are chance aligned with a background galaxy in the appropriate magnitude range.}
\tablenotetext{$\dagger$}{The probability that at least 1, 2, 3 or 4 cluster member sources are chance aligned with a background galaxy in the appropriate magnitude range.}
\end{deluxetable*}

As Figure \ref{fig:2color} shows, we have 7 stars in the "blue" region of the color-color plot, which we used to establish a selection rule
to clean the excess region of the plot of possible spurious detections. We only accepted stars as true excess stars that: 1) lie at least 
3$\sigma_{24}$ (their own $\sigma$ and not the average [24] error) from the trend line; 2) have
[24] data at least 3$\sigma_{24}$ from the best fitting SED solution also; and 3) are point sources on the images and
have no noise anomalies. All stars in the "blue" region failed these criteria. From the 19 stars that lie in the excess ("red") region of the 
color-color diagram, fifteen were eliminated as debris disk candidates for the following reasons. Only 8 were 3$\sigma_{24}$ 
from the trend line: 181, 143, 100, 77, 134, 188, 24, and 2 (in the nomenclature of Table \ref{tab:solutions} and Figure \ref{fig:2color}). 
Star 188 turned  out to be contaminated by a minor planet, which was identified by comparing 
scanlegs separately. Stars 24 and 2 are resolved doubles on the higher resolution 2MASS and SDSS images, so we excluded them from 
our list. 

Star 100 is contaminated by a faint background galaxy, which was visible as a faint nebulosity next to the star.
The probability for other sources of a 24 $\micron$ excess arising through a chance alignment with distant galaxies can be determined from galaxy counts \citep{papovich04}. 
Our $\sim 0.15$ mag [24] excess criterion results in different flux values identified as excesses as
a function of source brightness. We estimated the probability of chance alignments by dividing our sample into 1 magnitude bins and running a Monte Carlo code 
with the number of sources in the bin and the number of extragalactic sources corresponding to 0.15 magnitude excess value for the specific bin. The matching radius 
for the chance alignment was chosen to be $r=3.6''$ and the code was ran to 10000 simulations per magnitude bin. We summarize the simulation numbers with the 
probabilities of at least $n$ number of chance alignments in each bin in Table \ref{tab:chance}.
The probability that star 143 with $[24]=7.04$ mag is a chance alignment with a background galaxy is very low ($< 3$\%), so it is very likely to be a true debris disk
star. The probability that at least two sources (star 100 and 134) are contaminated by a background galaxy in the 8-9 magnitude bin is also very low ($< 4$\%), 
and since star 100 is already contaminated, we classify star 134 as a real debris disk star also. The likelihood
that stars 77 and 181 are contaminated within $3.6''$ is high ($\sim 90$\%). However, there is no indication of any positional
offset between K$_{S}$ and [24], even at the $1''$ level, so this likelihood is probably overestimated.

\begin{figure*}[t]
\begin{centering}
\includegraphics[angle=0,scale=1]{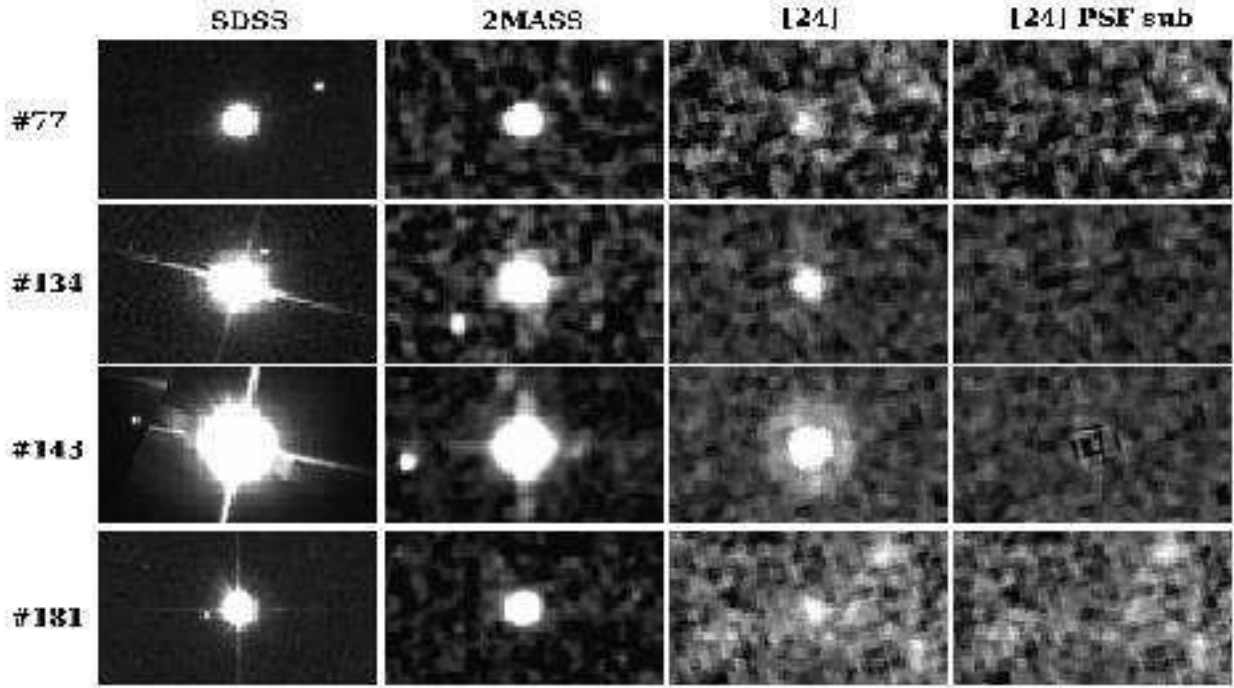}
\leavevmode
\caption{SDSS, 2MASS, 24 micron, and 24 micron PSF subtracted images for stars 77, 134, 143, and 181. The fields of view (FOV) for the images are 
$69.9'' \times 37.97''$ and they have linear flux scaling.}
\label{fig:exccand}
\end{centering}
\end{figure*}

We determine stars 143 and 134 to be definite debris disk stars in Praesepe and list stars 77 and 181 as possible debris disk stars. We
show these sources in Figure \ref{fig:exccand} and detail their properties in \S 4.4. Figure \ref{fig:exccand} shows that the fields are 
clean and that the sources are point-like. The PSFs were centered on the 24 $\micron$ sources with {\it IRAF}'s centroid algorithm. As Table \ref{tab:solutions}
shows, the coordinate center of the excesses is closer than $1''$ to the 2MASS coordinates for the debris disk candidates.

\subsection{The SED fit selection}

The 2MASS data are only useful in selecting debris disk candidates if the K$_{S}$ magnitude is
photospheric. Since the threshold of the K$_{S}$-[24] color above which a star is selected to be a debris disk candidate depends on the 
spectral type (the determination of which depends on correct r band photometry), we also fit the photometric data for all stars
within the trend curves with model spectra to look for excess candidates. To be considered a debris disk candidate
in this region required even stronger selection criteria then in the case of the "excess region stars." Stars were selected to be candidates from this
region if their [24] photometry was at least 3$\sigma_{24}$ from the fitted SED and if the star was 3$\sigma_{24}+10\%$ (0.1 mag) from the trend line in 
the color-color plot. The 10\% is an allowance for systematic errors. None of the stars within the trend curves passed these criteria.

\begin{figure*}[t]
\begin{center}
\includegraphics[angle=0,scale=1.3]{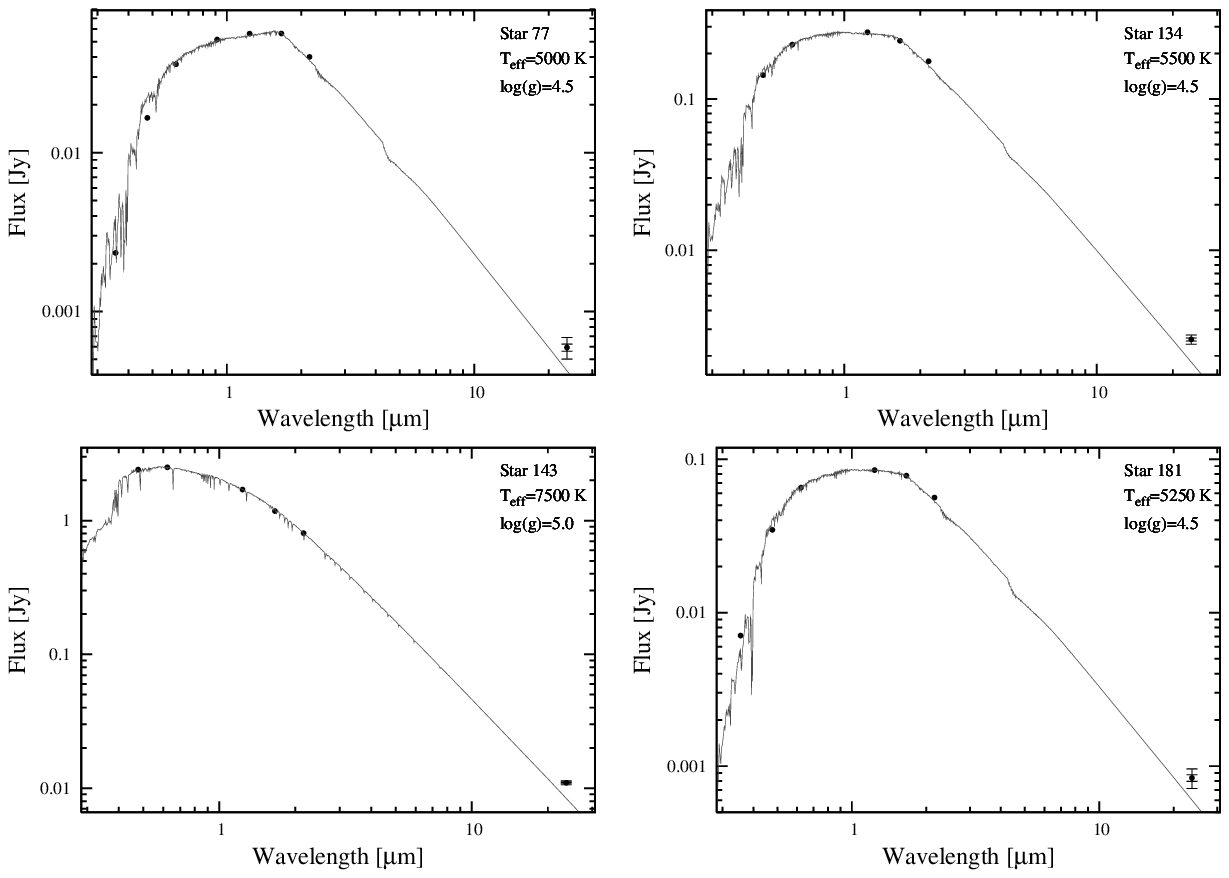}
\caption{The best fitting SEDs of the debris disk candidate stars with available optical, 2MASS, and [24] photometry. The [24] photometry is plotted
with 1 and 3$\sigma$ errors.}
\label{fig:sedplot}
\end{center}
\end{figure*}

\subsection{Praesepe white dwarfs}

We also checked whether any of the known eleven Praesepe white dwarfs \citep{dobbie06} were detected, indicating a possible white dwarf debris disk.
WD 0837+199 showed a strong signal in [24]. The UKIRT Infrared Deep Sky Survey (UKIDSS)
survey team (Sarah Casewell, private communication 2008) have found that this signal originates from a background galaxy a few arcseconds north of the WD.

\subsection{Debris Disk Candidates}

We discuss the four debris disk candidate stars in this section. None of these stars show extended emission (resolved disk), implying that the excess is confined to
the radius of the MIPS beam of $6''$ \citep{rieke04}, which is $\sim 1000$ AU at the distance of Praesepe. This is consistent with the sizes of already resolved 
mid-IR debris disks \citep{stapelfeldt04,su05,su08,backman09}. The best fitting SEDs of the debris disk candidate stars are plotted in Figure \ref{fig:sedplot}.

\subsubsection{Star \#77}

This star was identified on three separate scanlegs, with no contamination by minor planets. It is rather faint with $m_{\rm V}= 12.88$ mag. 
Its optical and NIR photometry was best fitted by the $T_{\rm eff}=5000~{\rm K}$ and $\log g=4.5$ (K3 V) Kurucz model \citep{kurucz11}.
\cite{franci03} used XMM-Newton to detect X-ray emission from it with a flux of $L_X=1.67\times10^{28}$ erg s$^{-1}$ in the {\it ROSAT} 0.1-2.4 keV band. 
They point out that the flux measured by ROSAT \citep{randich95} is a magnitude higher than theirs. The star is a cluster member cataloged 
in many papers \citep{wang95,kw,jones83}. 

\subsubsection{Star \#134}

Star \#134 (WJJP 179, KW 367) is a bright cluster member, with $m_{\rm V}=10.71$ mag. It was imaged on two scanlegs with high S/N. 
Its optical and NIR photometry was best fitted by the $T_{\rm eff}=5500~{\rm K}$ and $\log g=4.5$ (G8 V) Kurucz model
\citep{kurucz11} and we detect no extent to the stellar PSF core (Figure \ref{fig:exccand}). North of it by
$6''$, a fainter extended source is visible on both scanlegs. It has been found to be a triple system by \cite{mermilliod94} and the mass of the components was estimated by
\cite{hal03} using CORAVEL radial velocity measurements. The system consists of a wide pair, one of which is a spectroscopic binary with a period of 3.057 days. It is also
 a definite cluster member \citep{wang95,kw,jones83}. 
 
\subsubsection{Star \#143}

This is the brightest of all debris disk stars we observed, with $m_{\rm V}=8.04$ mag. 
Its optical and NIR photometry was best fitted with a $T_{\rm eff}=7500 ~{\rm K}$, $\log g=5.0$ (A7 V) Kurucz model \citep{kurucz11}.
The PSF subtraction was very clean, with no hint of any extended emission (Figure \ref{fig:exccand}). The star was discovered
to be a $\delta$ Scuti type of pulsating variable by \cite{paparo90} (HI Cnc, HD 73890, BD+19 2078). It has been
cataloged as a definite cluster member in many papers \citep{wang95,kw,jones83}. With high-resolution imaging surveys, \cite{mason93} found it to 
be a single star.

\subsubsection{Star \#181}

The star was identified on three separate scanlegs. The best fit to its photometry points was with a 
$T_{\rm eff}=5250 ~{\rm K}$, $\log g=4.5$ (K0 V) Kurucz model \citep{kurucz11}. 
It has been identified as a cluster member in many catalogs \citep{wang95,hambly95,kw,jones91}.
It was not identified as a close binary star in the surveys of \cite{bouvier01} and \cite{mermilliod99}. 
No extended emission is seen in our PSF subtracted image (Figure \ref{fig:exccand}). 

\section{Discussion}

We have found 4 sources out of 193 in the spectral range from A0 to K3 showing excess at 24 $\micron$.
One of our sources (star 143) is an A7 type star (out of 29 early-type stars), while the remaining three are G8, K0, and K3 (out of 164 solar-type stars), 
based on their photometric colors and fitted SEDs. Although the probability of chance alignments with faint
background galaxies within $3.6''$ are rather high for the K0 and K3 spectral-type sources, since the peaks of
their emission are well within $1''$ of the 2MASS coordinates they are likely excess sources. However, our statistics are
incomplete to their spectral limit. In our field of view there are 106 stars within F0 and G8 spectral-type, of which we 
detected 98, meaning we have an almost complete sample of sources within this spectral band. We use the excess fraction of 1/106 
for the solar-type star sample.

The excesses found around early type stars (B8-A9) are usually dealt with separately in the literature from the ones found around solar-type stars (F0-K4), because the 
dominant grain removal processes in the debris disks may not be the same and the 24 $\micron$ excesses probe significantly different distances
from the stars. These populations are also separated observationally, by the natural detection limits. 

In the following sections we analyze our results in the context of previous debris disk fractions observed around early- and solar-type stars. The errors on 
our debris disk fractions are given by Bayesian statistics detailed in the following \S 5.1. We contrast the results for early- and solar-type stars in 
\S 5.2 and \S 5.3 and discuss the implications for debris disk decay time scales in \S 5.4. In \S 5.5, we compare these results with a simple model
for the incidence of episodes like the LHB around other stars.

\subsection{Calulating errors on debris disk fractions}

Due to the small number of observations, we estimated our debris disk fractions and associated uncertainties using a Bayesian approach, which we
outline in this section.

If the fraction of objects with disks is $f_{\rm disk}$, derived from our observed number of disks ($n$) from a sample size of $N$, then
the posterior probability that $f_{\rm disk}$ has a certain value will be
\begin{equation}
P(f_{\rm disk}|n,N) \propto P(f_{\rm disk})P(n|f_{\rm disk},N).
\end{equation}
Here, $P(f_{\rm disk}|n,N)$ if the probability distribution for $f_{\rm disk}$, given that $n$ and $N$ are known. $P(f_{\rm disk})$ is
the prior distribution of $f_{\rm disk}$ and $P(n|f_{\rm disk},N)$ is the probability of observing that $n$ of $N$ sources have a disk, assuming
a certain value of $f_{\rm disk}$. $P(f_{\rm disk}|n,N)$ will be the posterior probability distribution for $f_{\rm disk}$ and $P(n|f_{\rm disk},N)$
is the likelihood function. If no prior assumption is made on the value of $f_{\rm disk}$, then the prior will be uniform, i.e.\ $P(f_{\rm disk})=1$. This
will be assumed, so that all information on $f_{\rm disk}$  originates from the data itself. The likelihood function, $P(n|f_{\rm disk},N)$, is a binomial
distribution, therefore
\begin{equation}
P(f_{\rm disk}|n,N) \propto f_{\rm disk}^{~~~n} (1-f_{\rm disk})^{N-n},
\end{equation}
where the binomial coefficient has been dropped because of its non-dependence on $f_{\rm disk}$, making it irrevelant in the posterior
distribution.

This equation is equivalent to a Beta ($B$) distribution with parameters $\alpha = n+1$ and $\beta = N - n +1$. The expectation value (posterior mean) 
of the $B$ distribution is simply
\begin{equation}
E(f_{\rm disk}) = \frac{\alpha}{\alpha+\beta} = \frac{n+1}{N+2}~,
\end{equation}
while its mode gives the regular ratio of $n/N$ (if $n>1$ and $N>2$). The 1$\sigma$ confidence region can be found
by integrating the central region that contains 68.3\% of the probability for the $B$ distribution. This was done in our paper by Monte
Carlo-type calculations. We simulated $10^7$ random variables from a $B$ distribution and searched for the bottom and upper limits
at the 15.85\% and 84.15\% percentiles.

We give our results with the expectation values and the upper and lower errors from the 1$\sigma$ limits. We decided to use expectation values
(posterior mean) over mode averages based on that our fractions are usually low making the distributions skewed. In such cases, they are better
described by their mean. For example, this will give an expected debris disk fraction of 
\begin{equation}
E(f_{\rm disk}) = \frac{1+1}{106+2} = 1.85 \%
\end{equation}
for our solar-type stars.

\subsection{The decay of the debris disk fraction in early-type stars}

A-type stars are well suited to search for excess emission originating from debris disks. The extended surveys of \cite{rieke05} 
and \cite{su06}, probed the excess fraction for A-type stars in the field and in associations between 
the ages of 5 and 850 Myr. Numerous observations have also determined the excess fraction for early-type stars in open clusters 
and associations \citep[e.g.][]{young04,gorlova04,gorlova06,siegler07,cieza08}.

\begin{figure*}[ht]
\begin{center}
\includegraphics[angle=0,scale=0.64]{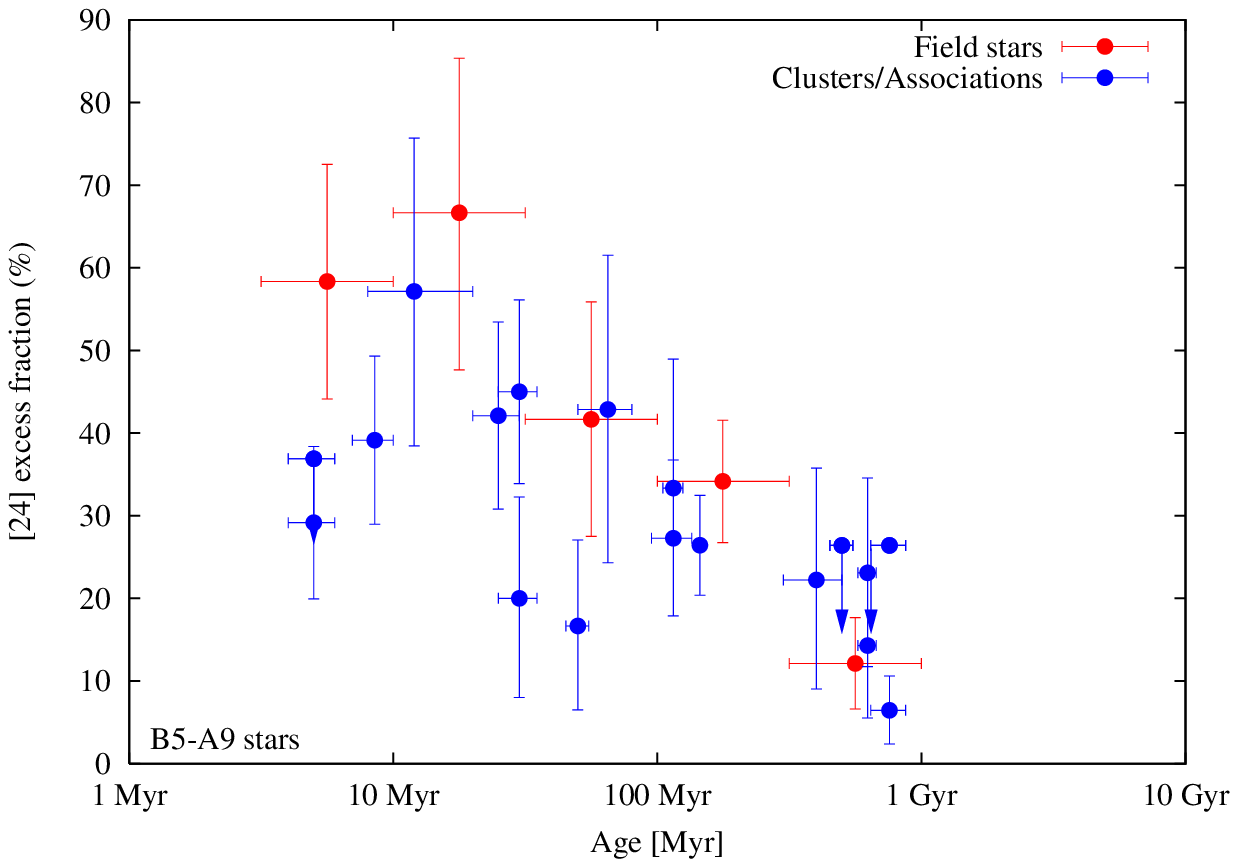}
\includegraphics[angle=0,scale=0.64]{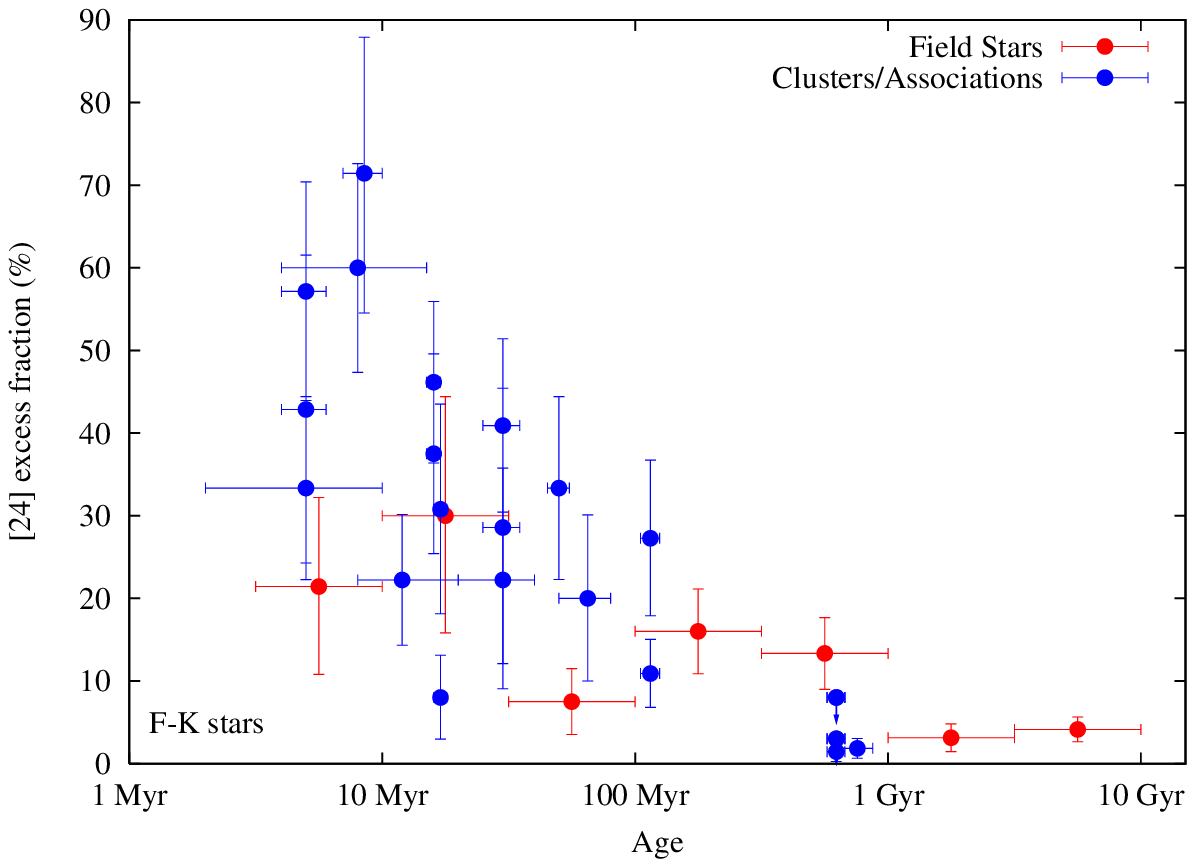}

\includegraphics[angle=0,scale=1.28]{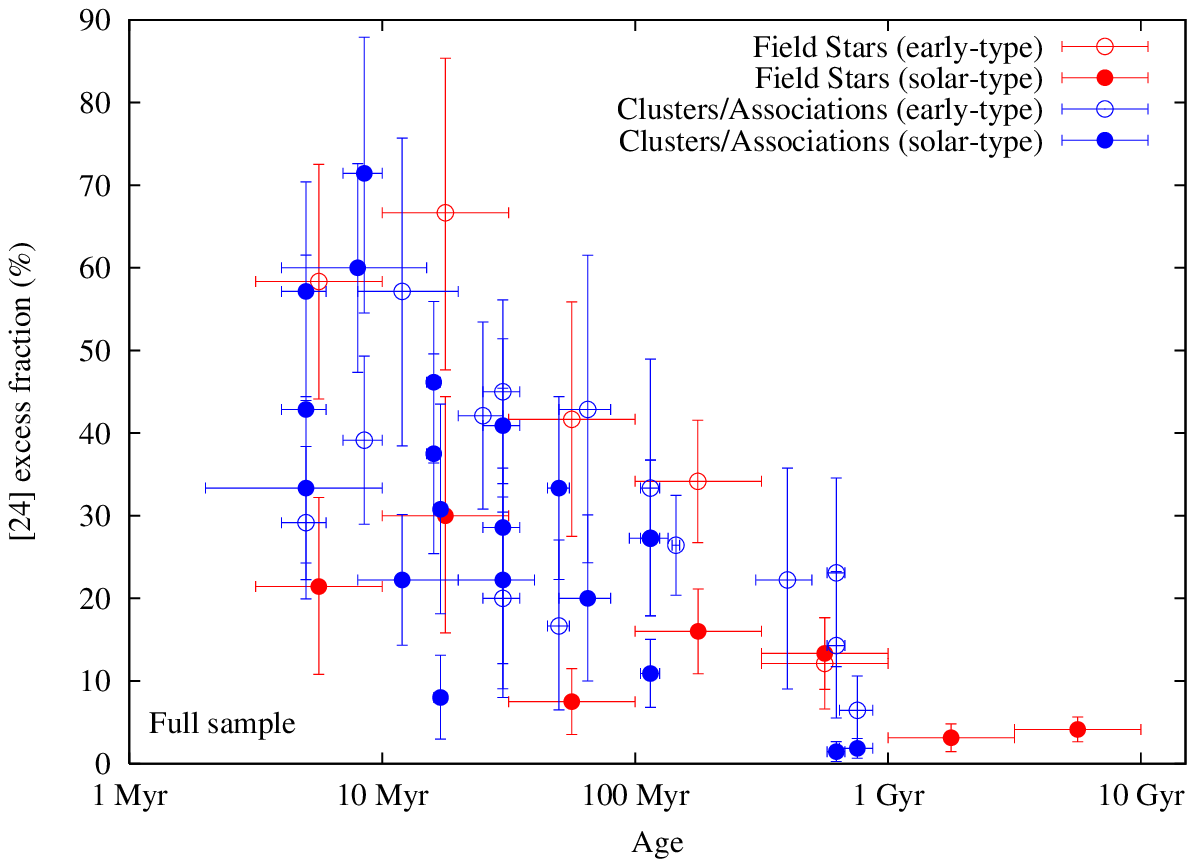}
\caption{{\it Top Left Panel:} The decay of the debris disk fraction for early type stars. {\it Top Right Panel:} The decay of the 
debris disk fraction for solar-type stars. {\it Bottom Panel:} The combined plot of all excess fractions.
The errors in excess fraction are the 1$\sigma$ errors from the beta distribution calculations (\S 5.1)
while the age errors are from the literature. The age "errors" for the field star sample show 
the age bins.
}
\label{fig:frac_all}
\end{center}
\end{figure*}

We compared our early spectral type excess fraction to the ones in the literature. We combined the data of \cite{rieke05}
and \cite{su06}, removing cluster and association members. Sources that were listed in both catalogs were adopted from \cite{su06}, due to
the improved reduction methods and photospheric model fits in the latter paper. Sources were counted as excess sources if their relative excess
exceeded 15\%. IRAS and ISO sources from the \cite{rieke05}
sample were removed, due to their higher -- 25\% -- excess thresholds. Our final age bins from the combined catalogs are listed in Table \ref{tab:gpk}.

\begin{deluxetable}{lll}
\tablewidth{0pt}
\tablecolumns{3}
\tablecaption{The field star sample excess ratios
at 24 $\micron$ at certain age bins for early type
stars \citep{rieke05,su06}.\label{tab:gpk}}
\tablehead{
\colhead{~~~~~~~~~~~~~~~~~~~~~Age~~~~~~~~~~~~~~~~~~~~~} & \multicolumn{2}{c}{Excess fraction} \\
\colhead{~~~~~~~~~~~~~~~~~~~~~[Myr]~~~~~~~~~~~~~~~~~~~~~} & \colhead{[\#]} & \colhead{[\%]}}
\startdata
3.16 	   - 10\dotfill 	& 6/10  & $58.3\pm 14.2$ \\
\phm{3.}10 - 31.6\dotfill  	& 3/4 	& $66.7^{+18.7}_{-19.1}$ \\
31.6       - 100\dotfill  	& 4/10  & $41.7\pm14.2$ \\
\phm{.}100 - 316\dotfill  	& 13/39 & $34.2\pm7.4$ \\
\phm{.}316 - 1000\dotfill 	& 3/31  & $12.1\pm5.5$ 
\enddata
\end{deluxetable}

We also compared our results to those from open cluster (and OB association) surveys by other groups. We list these clusters, their excess fraction,
age and the references for these parameters in Table \ref{tab:clusters}. The majority of these clusters are from MIPS group papers,
that used the same 15\% excess level threshold as we did in our study of Praesepe. The few others used similar thresholds, or as in the case of the
$\beta$ Pic MG study \citep{rebull08}, all excess sources that were identified exceeded their 20\% threshold with no sources between
15 and 20\%. We plot the excess fractions from all surveys with the field star samples in the {\it top left} panel of 
Figure \ref{fig:frac_all}.

\begin{deluxetable}{llllll}[]
\tablecolumns{6}
\tablewidth{0pt}
\tablecaption{The excess fraction at 24 $\micron$ for early type stars in clusters/associations.\label{tab:clusters}}
\tablehead{
\colhead{~~~~~Name~~~~~} & \colhead{Age} & \multicolumn{2}{c}{Excess fraction} & Exc.\  & Age \\
& \colhead{[Myr]} & \colhead{[\#]} & \colhead{[\%]} & \multicolumn{2}{c}{Reference~~~~~~}}
\startdata
Upper Sco\dotfill		& 5$\pm 1$		& 0/3   & 36.9$^{\ast}$			& 1,2	& 11 \\
Orion OB1b\dotfill		& 5$\pm 1$		& 6/22	& 29.2$\pm 9.2$			& 3	& 12 \\
Orion OB1a\dotfill		& 8.5$\pm 1.5$		& 8/21	& 39.1$\pm 10.2$		& 3	& 12 \\
$\beta$ Pic MG\dotfill		& 12$^{+8}_{-4}$	& 3/5	& 57.1$^{+18.6}_{-18.7}$	& 4	& 13 \\
Upper Cen\dotfill 		& 25$\pm 5$		& 7/17 	& 42.1$\pm 11.3$ 		& 1,2 	& 14\\
NGC 2547\dotfill 		& 30$\pm 5$		& 8/18 	& 45.0$\pm 11.1$ 		& 5,6  	& 5,6\\
IC 2602\dotfill			& 30$\pm 5$		& 1/8 	& 20.0$^{+12.3}_{-12.0}$	& 1   	& 15\\
IC 2391\dotfill 		& 50$\pm 5$ 		& 1/10 	& 16.7$^{+10.4}_{-10.1}$ 	& 7   	& 16\\
$\alpha$ Per\dotfill 		& 65$\pm 15$ 		& 2/5 	& 42.9$^{+18.7}_{-18.6}$ 	& 1,2	& 17,18\\
Pleiades\dotfill 		& 115$\pm 10$		& 2/7 	& 33.3$^{+15.6}_{-15.5}$ 	& 1	& 18,19,20\\
Pleiades\dotfill 		& 115$\pm 10$		& 5/20 	& 27.3$^{+9.5}_{-9.4}$ 		& 6 	& 18,19,20\\
NGC 2516\dotfill		& 145$\pm 5$		& 13/51 & 26.4$\pm6.0$			& 2	& 19,21\\
Ursa M\dotfill 			& 400$\pm 100$		& 1/7 	& 22.2$^{+13.5}_{-13.2}$ 	& 1   	& 22,23,24\\
Coma Ber\dotfill		& 500$\pm 50$		& 0/5	& 26.4$^{\ast}$			& 1,2   & 25 \\
Hyades\dotfill			& 625$\pm 50$ 		& 1/12 	& 14.3$^{+9.0}_{-8.8}$		& 1   	& 26 \\
Hyades\dotfill 			& 625$\pm 50$ 		& 2/11 	& 23.1$^{+11.5}_{-11.4}$	& 9   	& 26 \\
Praesepe\dotfill 		& 757$\pm 114$		& 1/29	&  6.5$\pm 4.1$			& 10   	& 10 \\
Praesepe\dotfill		& 757$\pm 114$		& 0/5	& 26.4$^{\ast}$			& 1	& 10
\enddata
\tablenotetext{$\ast$}{Upper limit}
\tablerefs{
(1) \cite{su06}; 
(2) \cite{rieke05}; 
(3) \cite{hernandez06};
(4) \cite{rebull08};
(5) \cite{young04};
(6) \cite{gorlova07}; 
(7) \cite{siegler07};
(8) \cite{gorlova06}; 
(9) \cite{cieza08};
(10) This work;
(11) \cite{preibisch02};
(12) \cite{briceno05};
(13) \cite{ortega02};
(14) \cite{fuchs06};
(15) \cite{stauffer97}; 
(16) \cite{barrado04};
(17) \cite{song01}; 
(18) \cite{martin01};
(19) \cite{meynet93}; 
(20) \cite{stauffer98}; 
(21) \cite{jeffries01};
(22) \cite{soderblom93}; 
(23) \cite{castellani02}; 
(24) \cite{king03}; 
(25) \cite{odenkirchen98};
(26) \cite{perryman98}}
\end{deluxetable}

The fifteen open clusters and associations follow the same trend as the field star sample, with the exception of IC 2602 \citep{su06} 
and IC 2391 \citep{siegler07}. Possible explanations for this deviation are explored in \cite{siegler07} and they conclude that the most
likely cause is the lack of a statistically large sample. The peak near $\sim 12$ Myr observed by \cite{currie08} is suggested.
Thereafter, the excess fraction shows a steady decline to the age of Praesepe ($\sim$ 750 Myr).
Although the single A7 debris disk star we observed is not a statistically high number, the sample of 29 stars it was drawn from is high enough to indicate a real 
lack of debris disks around early-type stars at $\sim$ 750 Myr.

\subsection{The decay of the debris disk fraction for solar-type stars}

Detailed studies of the frequency of debris disks as a function of system age are useful tools to characterize belts of planetesimals
and their collisions around solar-type stars. They provide important proxies for comparisons between the Solar System and exoplanetary systems
in terms of planetary system formation and evolution. For example, observations at 70 $\micron$ show that Kuiper-belt-like planetesimal 
systems around solar-type stars can be rather common \citep[$\sim 16$\%;][]{trilling08}\citep[$\sim 14$\%;][]{hillenbrand08}, but are not
necessarily accompanied by 24 $\micron$ excess, which would be indicative of terrestrial planet formation.

\begin{deluxetable}{lll}
\tablecolumns{3}
\tablewidth{0pt}
\tablecaption{The field star sample excess ratios at 24 $\micron$ for solar-type stars at certain age bins from the compiled sample of \cite{trilling08}, \cite{beichman06} 
and the FEPS collaboration \citep{carpenter08a,carpenter08b,meyer08}.\label{tab:solarfield}}
\tablehead{
\colhead{~~~~~~~~~~~~~~~~~~~~~~~~~Age~~~~~~~~~~~~~~~~~~~~~~~~~} & \multicolumn{2}{c}{Excess fraction} \\
 & \colhead{[\#]} & \colhead{[\%]}}
\startdata
3.16 	   - 10\phm{00.} Myr\dotfill 	& 2/12  & 21.4$^{+10.8}_{-10.6}$ \\
\phm{3.}10 - 31.6\phm{0} Myr\dotfill  	& 2/8   & 30.0$^{+14.4}_{-14.2}$ \\
31.6       - 100\phm{0.} Myr\dotfill  	& 2/38  &  7.5$\pm 4.0$   \\
\phm{.}100 - 316\phm{.0} Myr\dotfill  	& 7/48  & 16.0$\pm 5.1$  \\
\phm{.}316 - 1000\phm{.} Myr\dotfill 	& 7/58  & 13.3$\pm 4.3$  \\
\phm{36.}1 - 3.16\phm{0} Gyr\dotfill	& 2/94  &  3.1$\pm 1.7$   \\
3.16 	   - 10\phm{00.} Gyr\dotfill	& 6/167 &  4.1$\pm 1.5$
\enddata
\end{deluxetable}

\begin{deluxetable}{llllll}
\tablecolumns{6}
\tablewidth{0pt}
\tablecaption{The excess fraction in [24] for solar-type stars in clusters/associations.\label{tab:solarcluster}}
\tablehead{
\colhead{~~~~~Name~~~~~} & \colhead{Age} & \multicolumn{2}{c}{Excess fraction} & Exc.\  & Age \\
& \colhead{[Myr]} & \colhead{[\#]} & \colhead{[\%]} & \multicolumn{2}{c}{Reference~~~~~~}}
\startdata
Orion OB1b\dotfill	&  5$\pm 1$		& 7/12	& 57.1$\pm 13.2$		& 1	& 11 \\
Upper Sco\dotfill	&  5$\pm 1$		& 5/16  & 33.3$^{+11.1}_{-11.0}$	& 2     & 12\\
Upper Sco\dotfill	&  5$\pm 1$		& 2/5 	& 42.9$^{+18.7}_{-18.6}$	& 3 	& 12\\
$\eta$ Cha\dotfill	&  8$^{+7}_{-4}$	& 8/13	&   60$\pm 12.6$		& 4	& 13,14 \\
Orion OB1a\dotfill	&  9$\pm 2$		& 4/5	& 71.4$^{+16.4}_{-16.9}$	& 1	& 11 \\
$\beta$ Pic MG\dotfill	& 12$^{+8}_{-4}$	& 5/25	& 22.2$\pm 7.9$			& 5	& 15 \\
Lower Cen C\dotfill	& 16$\pm 1$		& 11/24	& 46.2$\pm 9.8$			& 3   	& 16\\
Lower Cen C\dotfill	& 16$\pm 1$		& 5/14	& 37.5$^{+12.1}_{-12.0}$	& 2   	& 16\\
Upper Cen L\dotfill	& 17$\pm 1$ 		& 3/11 	& 30.8$^{+12.7}_{-12.6}$	& 3   	& 12\\
Upper Cen L\dotfill	& 17$\pm 1$ 		& 1/23 	&  8.0$^{+5.1}_{-5.0}$		& 2   	& 12\\
NGC 2547\dotfill 	& 30$\pm 5$ 		& 8/20 	& 40.9$\pm 10.5$		& 6	& 6\\
Tuc-Hor\dotfill		& 30$\pm 5$		& 1/7	& 22.2$^{+13.5}_{-13.2}$	& 5	& 17\\
IC 2602\dotfill		& 30$\pm 5$		& 1/5	& 28.6$^{+16.9}_{-16.4}$	& 2	& 18\\
IC 2391\dotfill		& 50$\pm 5$ 		& 5/16  & 33.3$^{+11.1}_{-11.0}$	& 7   	& 19\\
$\alpha$ Per\dotfill 	& 65$\pm 15$ 		& 2/13 	& 20.0$^{+10.1}_{-10.0}$ 	& 2	& 20,21\\
Pleiades\dotfill 	& 115$\pm 10$		& 5/53 	& 10.9$\pm 4.1$			& 8 	& 21,22,23\\
Pleiades\dotfill 	& 115$\pm 10$		& 5/20 	& 27.3$\pm 9.4$			& 2 	& 21,22,23\\
Hyades\dotfill		& 625$\pm 50$ 		& 0/67 	&  2.7$^{\ast}$			& 9   	& 24\\
Hyades\dotfill 		& 625$\pm 50$ 		& 0/22 	&  7.7$^{\ast}$			& 2   	& 24\\
Praesepe\dotfill 	& 757$\pm 114$		& 1/106	&  1.9$\pm 1.2$			& 10  	& 10
\enddata
\tablenotetext{$\ast$}{Upper limit}
\tablerefs{
(1) \cite{hernandez06};
(2) \cite{carpenter08b};
(3) \cite{chen05}; 
(4) \cite{gautier08};
(5) \cite{rebull08};
(6) \cite{gorlova07};
(7) \cite{siegler07}; 
(8) \cite{gorlova06};  
(9) \cite{cieza08}; 
(10) This work;
(11) \cite{briceno05};
(12) \cite{preibisch02};
(13) \cite{mamajek99};
(14) \cite{lyo04};
(15) \cite{ortega02};
(16) \cite{mamajek02};
(17) \cite{rebull08}, with arbitrary errors adopted from similar age clusters;
(18) \cite{stauffer97};
(19) \cite{barrado04};
(20) \cite{song01}; 
(21) \cite{martin01};
(22) \cite{meynet93};
(23) \cite{stauffer98}; 
(24) \cite{perryman98}}
\end{deluxetable}

To provide a large sample, we merged the 24 $\micron$ data of \cite{trilling08}, \cite{beichman06} and that of the FEPS group
\citep{carpenter08a,carpenter08b,meyer08} resulting in a database of 425 solar-type field stars
with age estimates in the range from 3.16 Myr to 10 Gyr. The tables in \cite{trilling08} include the results of \cite{bryden06} and \cite{beichman05} 
with their photometry data reevaluated with the same procedures as the newer \cite{trilling08} sample. 
We divided this database into the same logarithmic age bins as we did for the early-type field star sample
and calculated the debris disk fraction in these bins using the 15\% threshold in excess emission at 24 $\micron$. The debris disk fractions are 
summarized in Table \ref{tab:solarfield}.

We also compiled results at 24 $\micron$ from the literature on debris disk fractions around solar-type stars in open clusters and associations.
They are summarized in Table \ref{tab:solarcluster}. The excess fractions for the combined sample of solar-type stars are plotted in the {\it top right} panel of 
Figure \ref{fig:frac_all}. The plots show a significantly larger scatter in the excess fractions for solar-type than for early-type stars. A second interesting
feature is a possible environmental effect on the fraction of debris disks around solar-type stars. Although not pronounced -- and possibly strongly effected
by sampling biases -- there seems to be higher fraction of debris disk stars in clusters/associations than in the field.

In Praesepe, the few debris disk candidate stars (from a statistically large sample of 106 stars) implies that 
the planetary systems in the 1-40 AU zones around solar-type stars have generally reached a quiescent phase. This behavior can be compared with that of the field star 
sample, which levels off at a few percent at ages $>1$ Gyr. This result may seem surprising given the LHB period of the Solar System, but it is actually consistent with 
the models of \cite{gomes05} and \cite{thommes08}. The LHB was modeled in these papers to be a result of instability in the planetary system, caused by either strong 
interaction at the mean motion resonances of Jupiter and Saturn or that of Uranus and Neptune. In both cases the outer planetary disk is destabilized, causing planetesimals 
to migrate inward and initiate a collisional cascade. The models of \cite{gomes05} show a wide range of ages (192 Myr -- 1.1 Gyr) when the LHB can occur, but they are more 
likely to be initiated at the earlier ages. The timing of the cascade depends on a few initial conditions that can be set to realistic parameters to give any of the solutions. 
The paper by \cite{strom05} also agrees that the LHB was a catastrophic event, lasting between 10 and 150 Myr, however they argue that the characteristics of the craters 
found on the inner planets originating from that epoch are more likely to be from main belt asteroids.\footnote{\cite{hartmann00} and \cite{morbidelli01} argued that the LHB was the 
tail end of a monotonically decreasing impactor population. This theory was questioned by \cite{bottke07}, who computed the probability of the cratering records 
being created by it, and could rule it out at a 99.7\% (3$\sigma$) confidence level.} The collisional cascade or "terminal cataclysm" model is also supported by recent
studies of Hadean-era zircons on Earth \citep{trail07}.

\subsection{Evolutionary differences between the debris disks around early- and solar-type stars}

To illustrate the differences between the evolution of debris disks around early- and solar-type stars, we combined the top panel plots in Figure \ref{fig:frac_all}
in the bottom panel of the same figure. There appears to be an upper envelope to the excess fraction as a function of age, as if there
were a theoretical maximum number of debris disks possible at any age. There is substantial scatter below this envelope.

\begin{figure*}[t]
\begin{center}
\includegraphics[angle=0,scale=1.02]{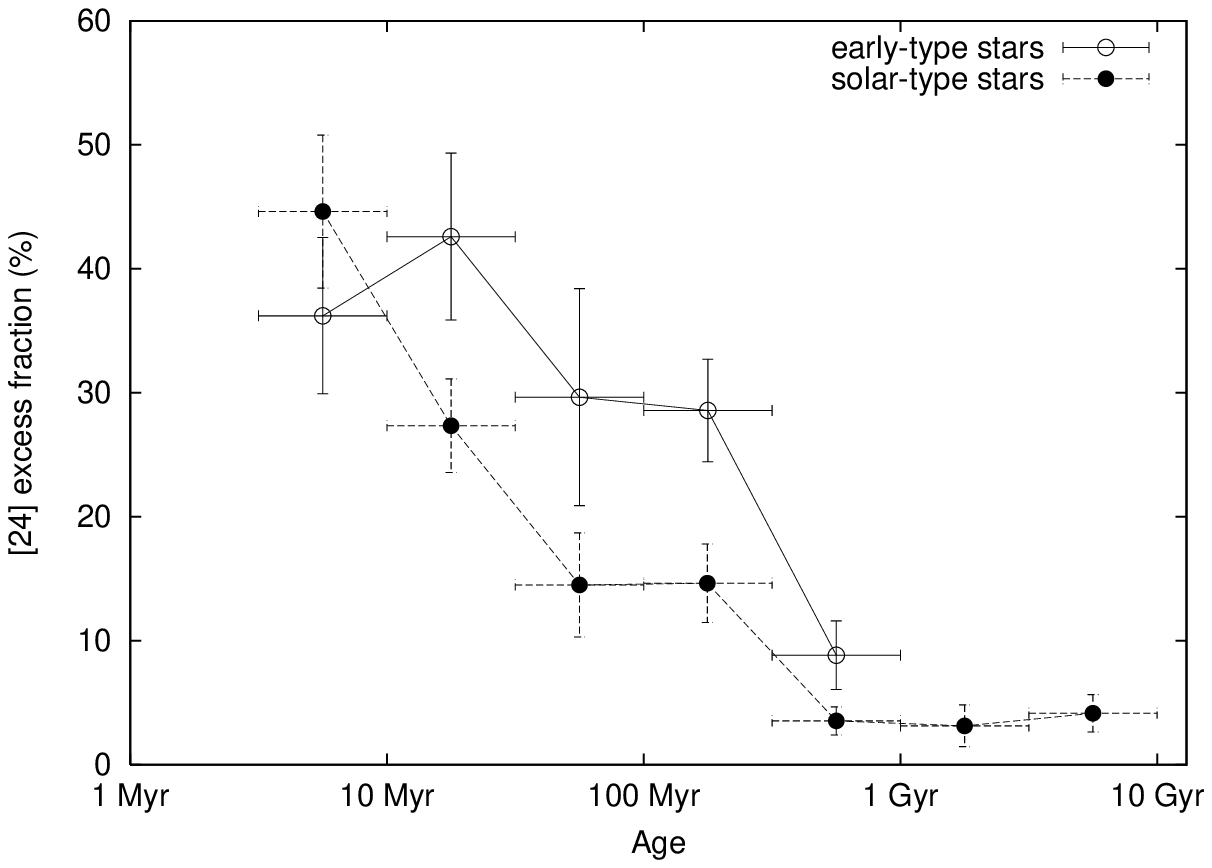}
\includegraphics[angle=0,scale=1.02]{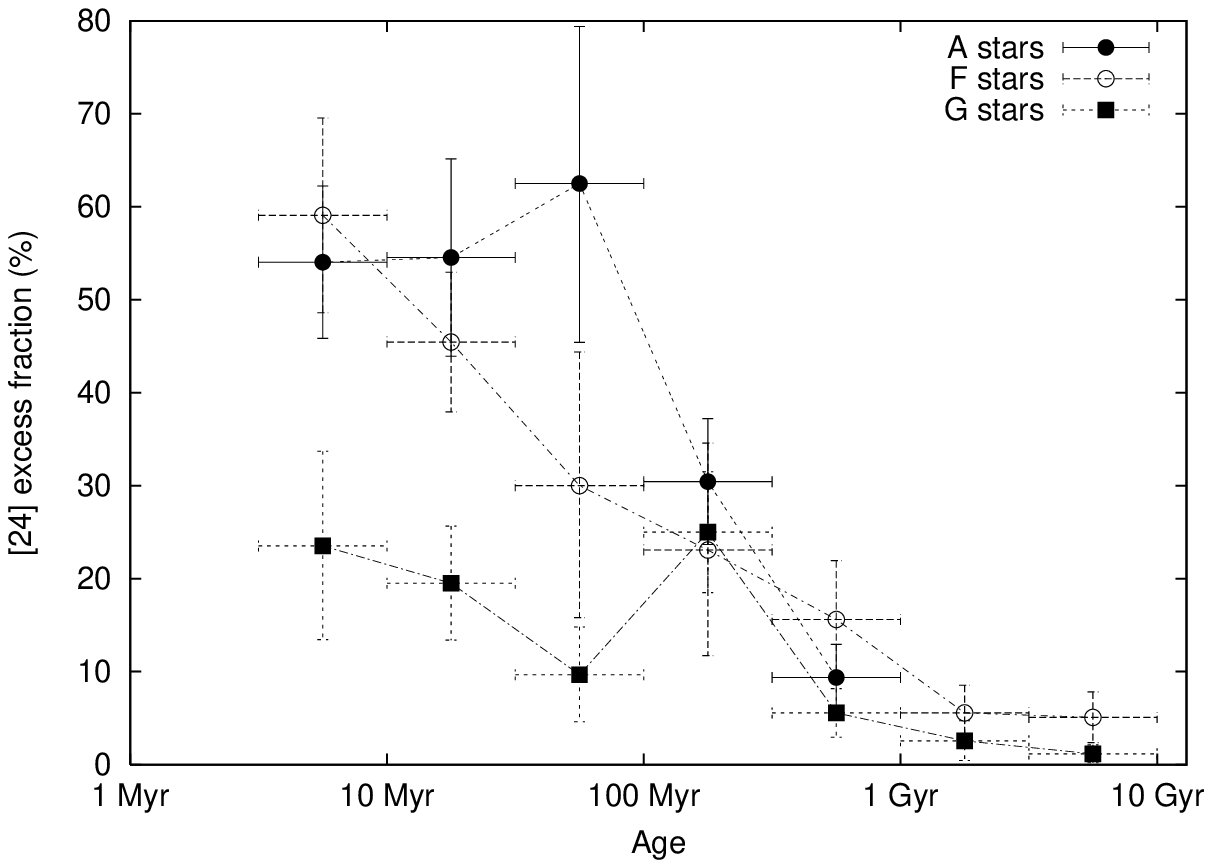}
\caption{
{\it Top panel:} The difference between the decaying trend for early- and solar-type stars, in a binned data plot. 
{\it Bottom panel:} The difference between the decaying trend for A, F and G spectral-type stars, in a binned data plot. 
The errors in the excess fraction are from beta distribution calculations (\S 5.1), while the "error bars" in the ages show the age bins.
The numerical values for the data points are summarized in Table \ref{tab:rebin}.
}
\label{fig:rebin}
\end{center}
\end{figure*}

Figure \ref{fig:frac_all} shows that there is a subtle difference between the evolution of debris disks around early- and solar-type stars.
To reduce the effects of observational biases (such as detection thresholds) and sampling differences (number of stars in clusters), we rebinned all the data to a more 
homogeneous sampling.
We used the same logarithmic age bins as we did for the field star samples: 3.16-10, 10-31.6, 31.6-100, 100-316 Myr, and 0.316-1, 
1-3.16, and 3.16-10 Gyr. The result is shown in Figure \ref{fig:rebin}, along with a second plot that shows the decay trends for A, F and G spectral-type stars separately. 
The data for all rebinned decay trends are summarized in Table \ref{tab:rebin}. The "rise-and-fall" characteristics
for early-type stars is confirmed \citep{currie08}, but with a quick drop-off at later ages. The solar-type stars show a monotonic 
decaying trend that reaches a constant of a few percent at later ages. The most important feature though is that the trends have different timescales.

\begin{deluxetable*}{lllllllllll}
\tablecolumns{11}
\tabletypesize{\footnotesize}
\tablewidth{0pt}
\tablecaption{The percent of debris disks in a rebinned distribution,
as a function of stellar spectral-type. \label{tab:rebin}}
\tablehead{
\colhead{Age}	& \multicolumn{2}{c}{Early-type stars} & \multicolumn{2}{c}{Solar-type stars}  & \multicolumn{2}{c}{A-type stars}    & \multicolumn{2}{c}{F-type stars}    & \multicolumn{2}{c}{G-type stars}   \\ 
		& \multicolumn{2}{c}{Excess fraction}  & \multicolumn{2}{c}{Excess fraction}  & \multicolumn{2}{c}{Excess fraction} & \multicolumn{2}{c}{Excess fraction} & \multicolumn{2}{c}{Excess fraction}\\
 \colhead{}	& \colhead{[\#]} & \colhead{[\%]}      & \colhead{[\#]} & \colhead{[\%]}      & \colhead{[\#]} & \colhead{[\%]}	    & \colhead{[\#]} & \colhead{[\%]}	  & \colhead{[\#]} & \colhead{[\%]}}
\startdata
3.16 	   - 10\phm{00.}Myr  & 20/56  & 36.2$\pm 6.3$	      & 28/63  & 44.6$\pm 6.2$ & 19/35 & 54.05$\pm  8.2$ 	 & 12/20 & 59.09$\pm 10.5$	   & 3/15  & 23.5$^{+10.1}_{-10.2}$ \\
\phm{3.}10 - 31.6\phm{0}Myr  & 22/52  & 42.6$\pm 6.7$	      & 37/137 & 27.3$\pm 3.8$ & 11/20 & 54.55$\pm 10.6$ 	 & 19/42 & 45.45$\pm 7.5$	   & 7/39  & 19.5$\pm 6.1$ \\
31.6       - 100\phm{0.}Myr  & 7/25   & 29.6$^{+8.8}_{-8.7}$    & 9/67   & 14.5$\pm 4.2$ & 4/6   & 62.50$^{+17.1}_{-16.9}$ & 2/8	 & 30.00$^{+14.2}_{-14.4}$ & 2/29  &  9.7$\pm 5.1$ \\
\phm{.}100 - 316\phm{.0}Myr  & 33/117 & 28.6$\pm 4.1$	      & 17/121 & 14.6$\pm 3.2$ & 13/44 & 30.43$\pm 6.8$ 	 & 2/11  & 23.08$^{+11.4}_{-11.5}$ & 10/42 & 25.0$\pm 6.5$ \\
\phm{.}316 - 1000\phm{.}Myr  & 8/100  &  8.8$\pm 2.8$	      & 8/253  &  3.5$\pm 1.1$ & 5/62  &  9.38$\pm 3.6$ 	 & 4/30  & 15.62$\pm 6.3$	   & 3/70  &  5.6$\pm 2.6$ \\
\phm{36.}1 - 3.16\phm{0}Gyr  & -      &  -		      & 2/94   &  3.1$\pm 1.7$ & -     & -      		 & 2/52  &  5.56$\pm 3.0$	   & 0/37  &  2.6$^{+2.1}_{-2.2}$ \\
3.16 	   - 10\phm{00.}Gyr  & -      &  -		      & 6/167  &  4.1$\pm 1.5$ & -     & -      		 & 2/57  &  5.08$\pm 2.7$ 	   & 0/85  &  1.2$\pm 1.0$ 
\enddata
\end{deluxetable*}

The fraction of infrared excesses at a given age range is set by the interplay of the occurrence rate of the collisional 
cascades for each system, the longevity of the dust produced in these cascades, and our ability to detect the debris at 
the distance of the given cluster. Detailed modeling of these processes is required to interpret the different rate of 
decline in the debris disk fraction between early- and solar-type stars. Although such modeling is beyond the scope of 
this paper, three possible explanations can be invoked to explain qualitatively the faster decline of excess fraction 
around solar-type stars. First, the dust must be in the 24 $\micron$ emitting regions and
solar-type stars have about 50$\times$ smaller disk surface area in which a collisional cascade can produce warm enough dust. Second, 
the orbital velocity of planetesimals in the 24 $\micron$ emitting zone will be higher around solar-type than the early-type stars, possibly
accelerating the evolution of their debris disks. Third, the dust size distributions and lifetimes are different for the two groups of stars. 
These issues will be discussed in a forthcoming paper (G\'asp\'ar et al. 2009).

\subsection{Our results in context with the Late Heavy Bombardment}

The cratering record of all non geologically active rocky planets and moons in the inner Solar System reveal a period of very intense past bombardment.
Geochronology of the lunar cratering record shows that this bombardment ended abruptly at $\sim 700$ Myr \citep[see e.g.][]{tera73,tera74,chapman07}, 
but the scarcity of the lunar rock record prior to this event hinders accurate assessment of the temporal evolution of the impact rates or the length 
of the bombardment period. Dynamical simulations of different possible impactor populations show that an unrealistically massive impactor population 
would be required to maintain the impact rate measured at the end of the bombardment for a prolonged period, thus convincingly arguing for the bombardment 
being a short-duration spike in the impact rate \citep{bottke07}.
A possible explanation for this is that a dynamical instability initiated by the migration of the giant planets caused minor 
planetary bodies to migrate inwards from the outer region of the Solar System, bombarding the inner planets.
Modeling shows that this scenario can occur over a wide range of ages \citep{gomes05}. \cite{strom05} show that it is possible instead that 
main belt asteroids bombarded the planetary system.

\begin{figure}[t]
\begin{center}
\includegraphics[angle=0,scale=1.1]{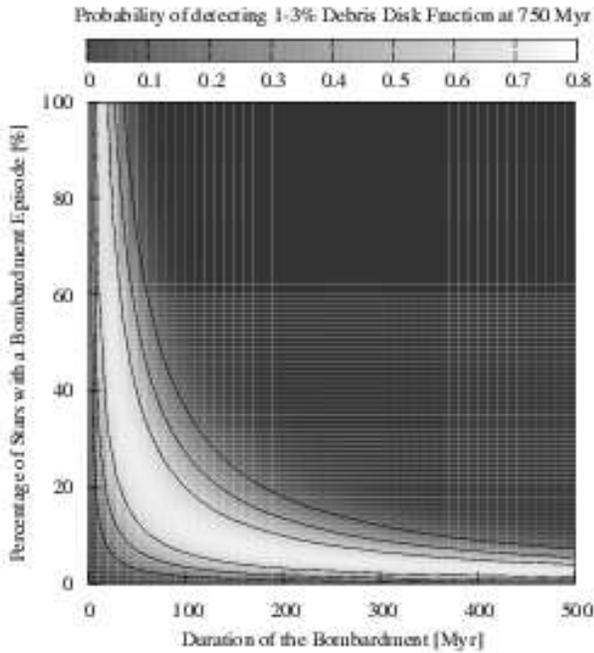}
\caption{The probability of detecting the 1-3\% (2/(106+29)) debris disk fraction observed at Praesepe, as a function of the percentage of stars that undergo LHB type 
debris disk generation and the duration of the events. The contour lines are at 20, 40 and 60\% probability.}
\label{fig:lhb}
\end{center}
\end{figure}

We performed a Monte Carlo simulation to evaluate our observed debris disk fraction in the context of the evidence from the LHB. Our goal was to constrain the fraction
of the solar-type stars that undergo LHB (or fine dust generation) and the duration of these events. We presumed in our models that all LHB events could be
detected in the existing debris disk surveys and that they had an equal probability of occurring once from 100 Myr to 1 Gyr. Both of these are strong assumptions. 
There is significant uncertainty on how much dust was generated and under what time scales during the LHB, making it difficult to relate the LHB unambiguously to
debris disks. However, given that {\it Spitzer} measurements of 24 $\micron$ excess emission are typically sensitive to a collisional cascade involving mass on the
order of a few lunar masses, and that such an episode has clearing time scale $\gtrsim 2$ Myr \citep{grogan01}, it seems plausible that the destruction of a few large 
asteroids can be detected in most observed systems. In our code we modeled clusters with 135 (106+29) members in 20000 simulations.
We varied the overall percentage of stars that will ever generate a debris disk from 0 to 100 \% and the duration of their bombardment episodes from 0 to 500 Myr. 
If the number of disks at 750 Myr were within our measured excess fraction of 1-3\%, the simulation was tagged as being consistent with our measurements, else it was 
tagged inconsistent. The overall probability of a given parameter pair is given by dividing the number of consistent simulations at a certain total disk fraction and 
duration timescale by the number of simulations (20000).

Our calculated probability map is shown in Figure \ref{fig:lhb}. The plot shows that the results are degenerate in the parameter space of 
d$t$ and $p_{\rm d}$, with d$t$ being the duration of a bombardment episode and $p_{\rm d}$ the percentage of stars to ever undergo such an event.
 Between the extremes of a very large percentage of the stars undergoing debris disk generation, but with a very short lifetime ($\sim$ 5-10 Myr) and a very 
small percentage ($< 5$ \%), with a long ($> 300$ Myr) lifetime there is a continuous set of solutions.

Our simple model allows the quantitative assessment of the probability of different types of LHB-like episodes. For example, we can exclude at a 3 \% significance 
level that 60\% of the stars undergo major orbital rearrangements, if this leads to debris production over 100 Myr. Similarly, very short debris producing events are unlikely,
because they would not produce observable disks, inconsistent with our results.

If we seek to evaluate the probability of strictly LHB-like debris producing episodes we can fix the length of the episode to 75 Myr, consistent with the duration estimated
for the inner Solar System and the other timescales discussed in \S 5.3. In this case our results show that up to 15-30 \% of the stars should undergo 
such a major orbital reorientation during the first Gyr of their evolution to be consistent with our modeling.


\section{Summary}

We conducted a 24 $\micron$ photometric survey for debris disks in the nearby ($\sim$ 180 pc) relatively old (750 Myr) Praesepe open cluster. 
The combined sample of SDSS, Webda, and 2MASS gave us a robust highly probable cluster member list. With simultaneous 
fitting of cluster distance and age we derived a series of solutions for both parameters as a function of metallicity (see Appendix). Our derived age for Praesepe is
757 Myr ($\pm$ 114 Myr at 3$\sigma$ confidence) and a distance of 179 pc ($\pm$ 6 pc at 3$\sigma$ confidence), not allowing for systematic errors.

Out of the 193 cluster members that we detected at all wavelengths in the combined catalog, 29 were early (B5-A9) and 164 later (F0-M0) spectral 
types. We found one star in the early and three in the later spectral type groups that show excess emission.
Up to near our completeness limit, with one debris disk star, there are 106 sources in the later spectral-type sample.
This result shows that only $6.5\pm 4.1$\% of early- and $1.9 \pm 1.2$\% of solar-type stars are likely to possess debris disks in the 1-40 AU 
zones. These values are similar to that found for old ($> 1$ Gyr) field stars. 

We place our results in context with the Late Heavy Bombardment theory of the Solar System. With simple Monte Carlo modeling we show that our 
observations are consistent with 15-30\% of the stars undergoing a major re-arrangement of the planetary orbits and a subsequent LHB-like episode once in their lifetime, 
with a duration period of 50-100 Myr.

We also summarize the results in the literature on the decay timescales of debris disks around early- and solar-type stars. We find that the decay timescale
for solar-type stars is shorter than for earlier-type stars. 

\acknowledgments

We would like to thank Sarah Casewell from the UKIDSS team for identifying the source of excess at WD 0837+199
to be a background galaxy. Our work is based on observations with \textit{Spit\-zer Space Telescope}, which is operated by the Jet Propulsion 
Laboratory, California Institute of Technology under NASA contract 1407. Support for this work was provided by NASA through Contract 
Number 1255094 issued by JPL/Caltech. This research made use of the SIMBAD database, operated at CDS, Strasbourg, France. Optical data
was in part obtained from the SDSS 5$^{\rm th}$ data release. This research has made use of the WEBDA database, operated at the Institute for Astronomy 
of the University of Vienna and data products from the Two Micron All Sky Survey, which is a joint project of the University of Massachusetts and the 
Infrared Processing and Analysis Center/California Institute of Technology, funded by the National Aeronautics and Space Administration and the National 
Science Foundation.
{\it Facilities:} \facility{{\it Spitzer} (MIPS)}

\appendix

\section{Age and distance estimate}

The precise value of the cluster age is important in constraining the debris disk fraction as a function of stellar age. The age and distance of Praesepe
have been a matter of debate, especially since it is an important step in the galactic distance ladder. The estimated
ages spread from $\log t=8.6$ all the way to $\log t=9.15$ (400 Myr -- 1.42 Gyr)\footnote{\cite{allen73,vandenberg84,tsvetkov93,gonzales06}}. 
Most papers list it as a coeval cluster with the Hyades because of their similar metallicities and spatial motions \citep[see, e.g.,][]{barrado98}.
The Hyades on the other hand has a better defined age of $\log t\approx8.8$ (625$\pm$50 Myr)\citep{perryman98,lebreton01}. If the 
clusters are coeval, their ages should agree within close limits.

Aside from using pulsating variables \citep{tsvetkov93} or stellar rotation \citep{pace04} to estimate the age of the cluster, the only method is to fit 
theoretical stellar evolution turnoff points on the observed CMD. This procedure involves a precise 
simultaneous fitting of the cluster distance, reddening, metallicity and age. 

\begin{figure*}[b]
\begin{center}
\includegraphics[angle=0,scale=0.56]{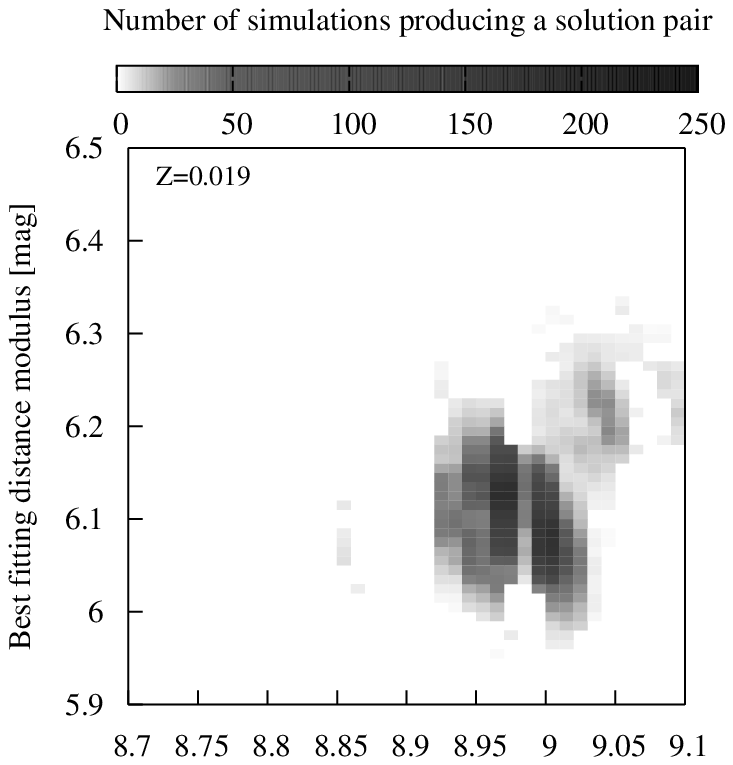}
\includegraphics[angle=0,scale=0.56]{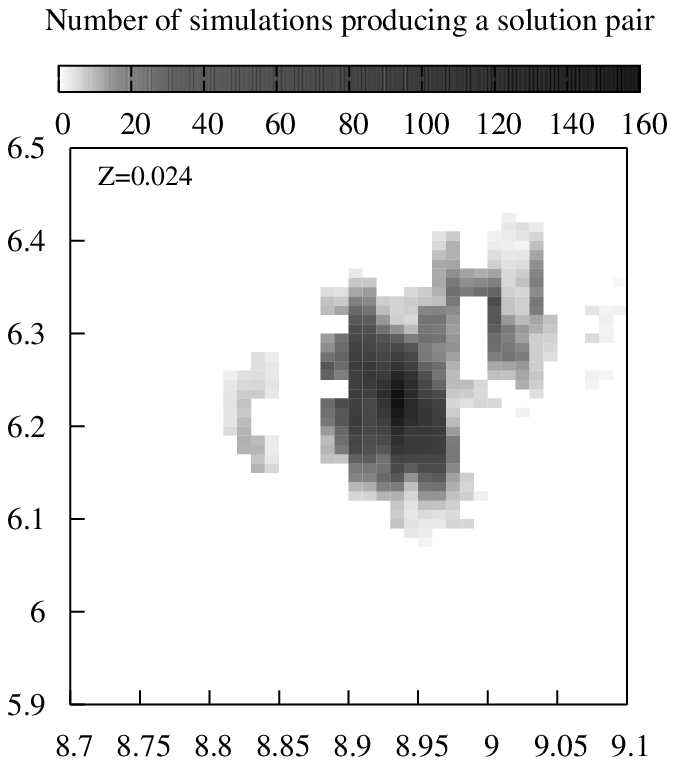}
\includegraphics[angle=0,scale=0.56]{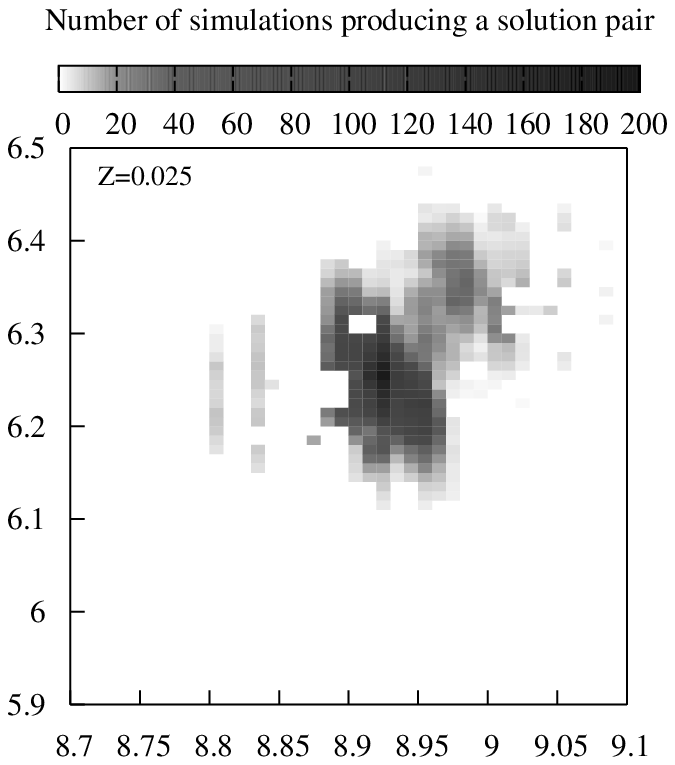}
\includegraphics[angle=0,scale=0.56]{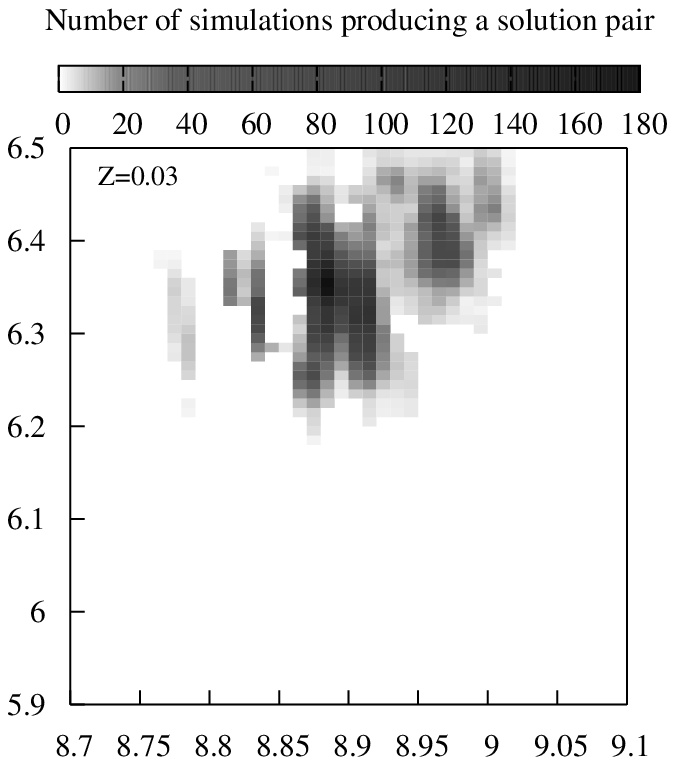}
\includegraphics[angle=0,scale=0.56]{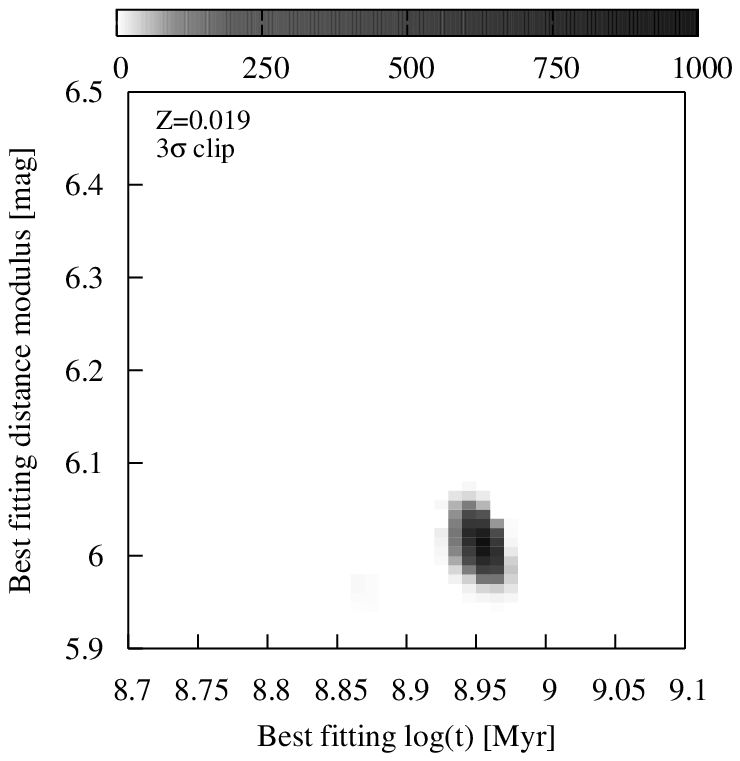}
\includegraphics[angle=0,scale=0.56]{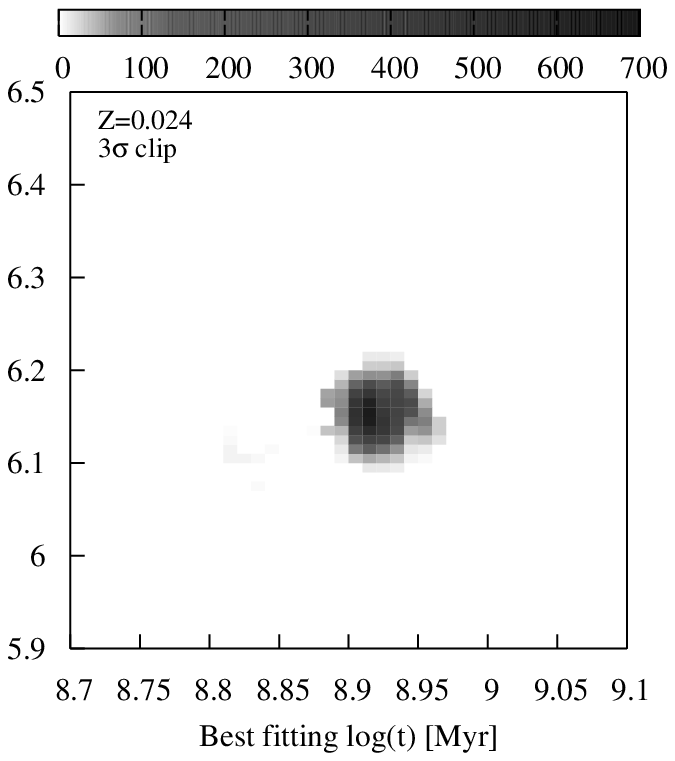}
\includegraphics[angle=0,scale=0.56]{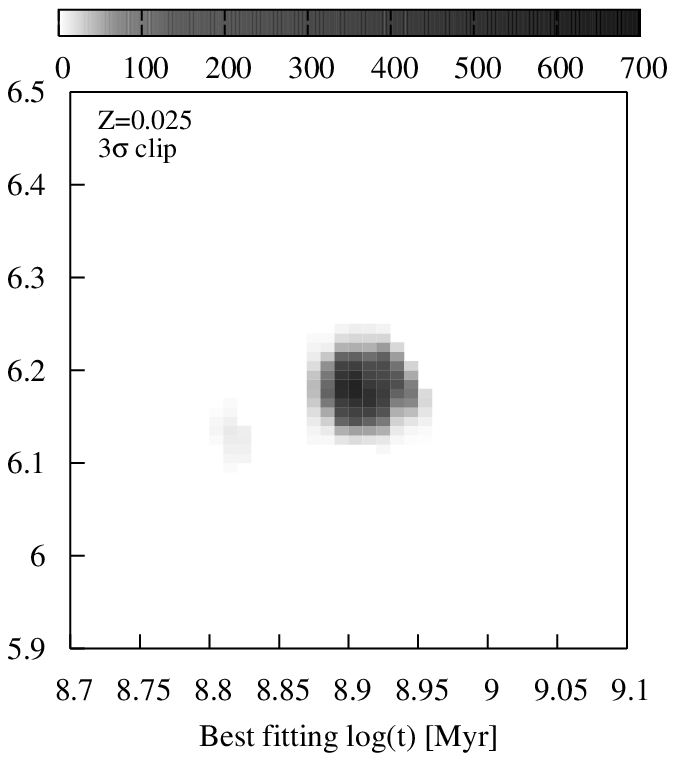}
\includegraphics[angle=0,scale=0.56]{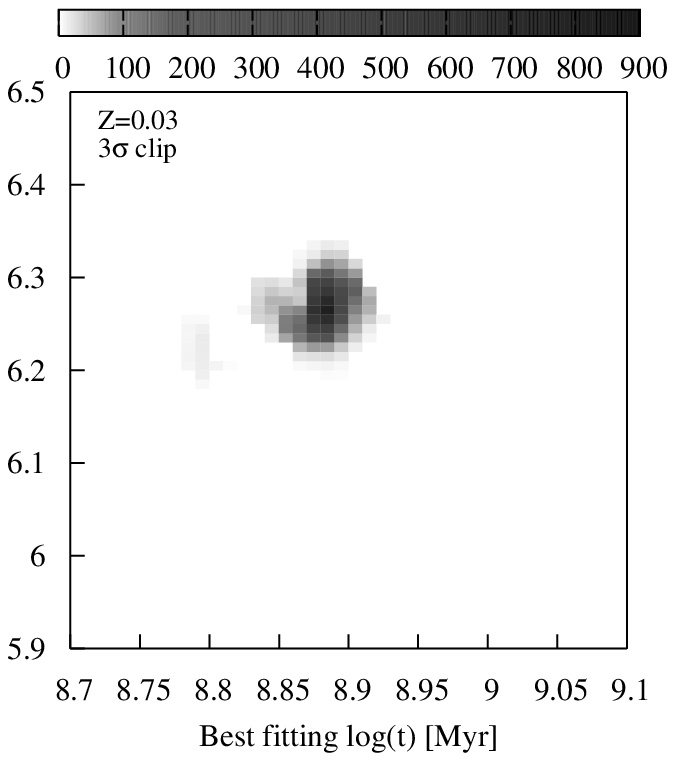}
\caption{The 2D probability maps show the number of solutions that were given for certain solution pairs
by the Monte Carlo isochrone fitting algorithm. The top row shows the fitting for the full sample, while
the bottom row gives the solutions after a 3$\sigma$ clipping iterational step. The fitted metallicities
are $Z=0.019$, 0.024, 0.025 and 0.03.}
\label{fig:isofit}
\end{center}
\end{figure*}

\begin{figure}[t]
\begin{center}
\includegraphics[angle=0,scale=0.93]{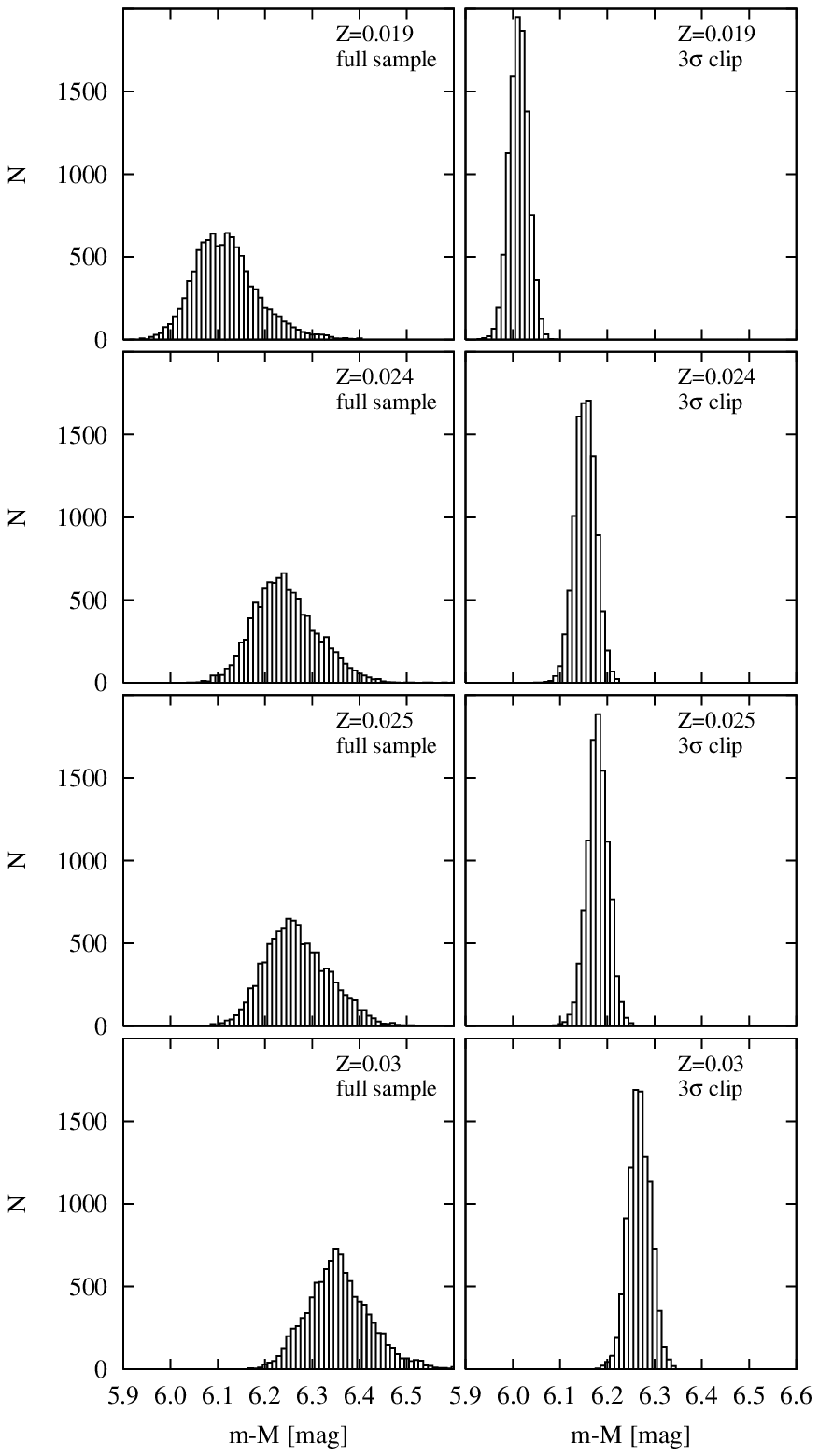}
\includegraphics[angle=0,scale=0.93]{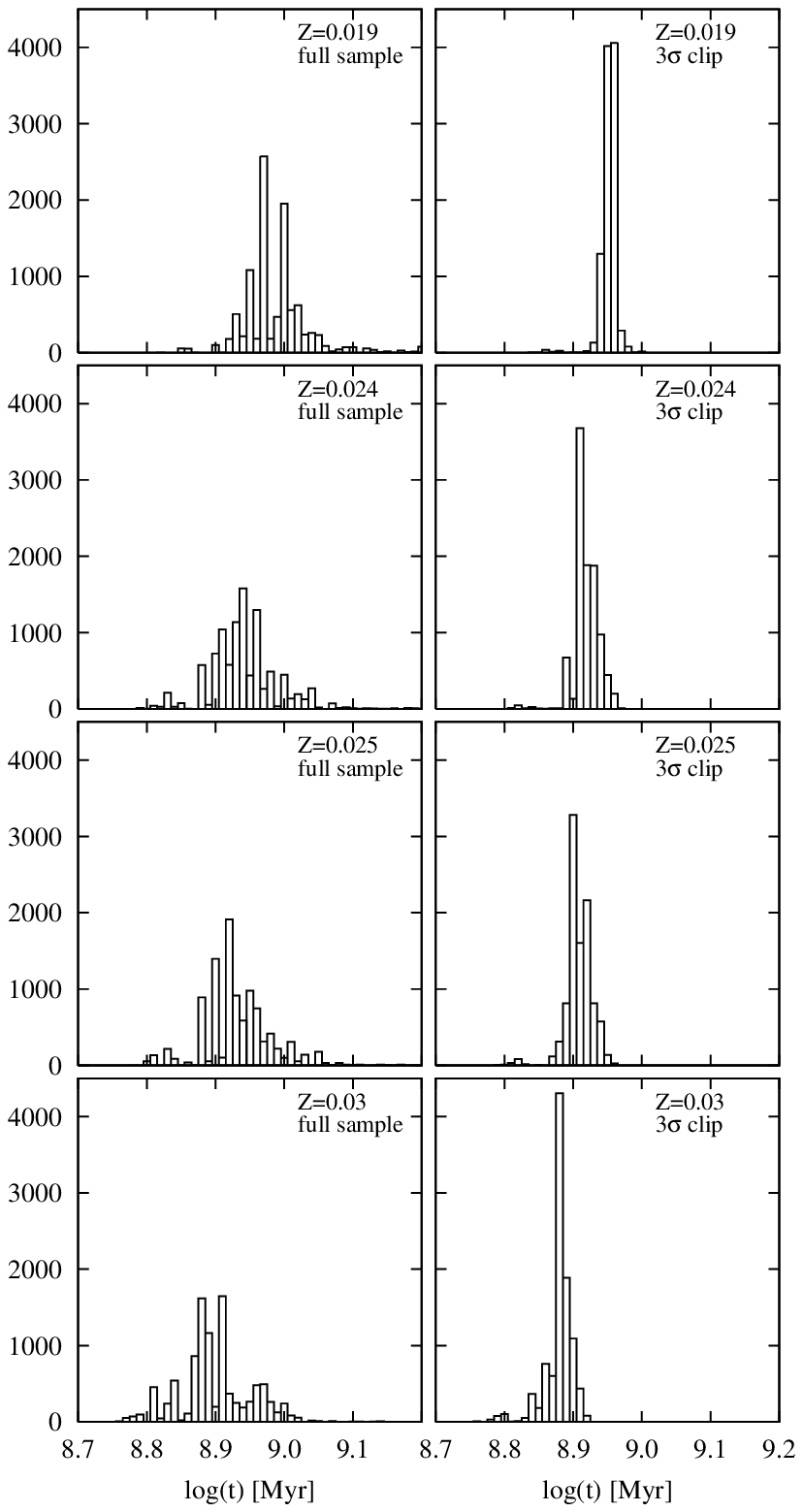}
\caption{These plots show the 1D representation of Figure \ref{fig:isofit}, separately
for $m-M$ and $\log(t)$. The distributions get much narrower after the 3$\sigma$ clipping 
iterations. These plots clearly show that as you go to more metal rich isochrones, the best
fitting isochrones will be younger and more distant.}
\label{fig:histos}
\end{center}
\end{figure}

\begin{figure}[]
\begin{center}
\includegraphics[angle=0,scale=1.2]{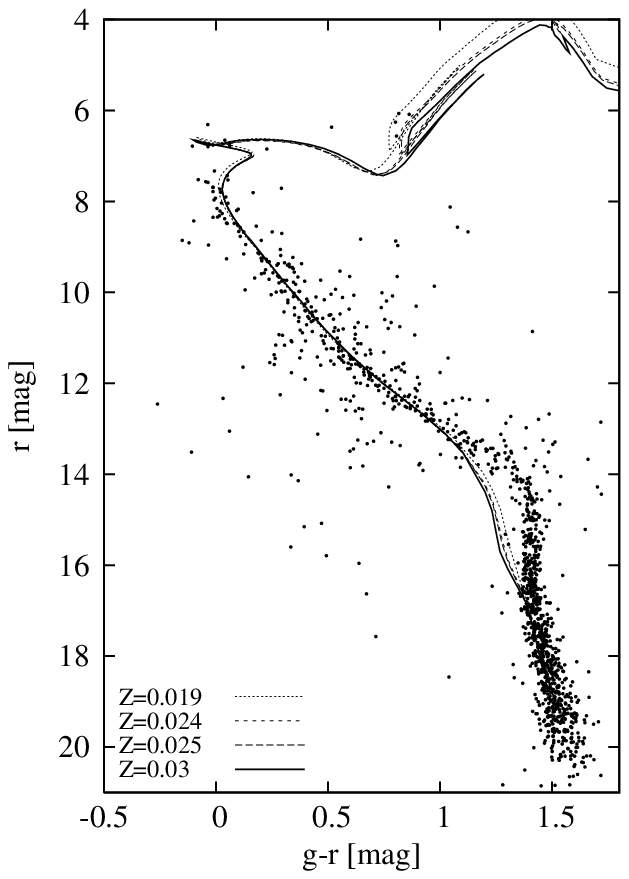}
\caption{The best fitting isochrones for all metallicities in the literature plus solar. The dotted line isochrone that deviates from the rest at high luminosities is
the solar (Z=0.019) isochrone.}
\label{fig:grr}
\end{center}
\end{figure} 

The metallicity of Praesepe has been revisited many times. The value of \cite{boesgaard88} of $[{\rm Fe}/{\rm H}] = 0.13 \pm 0.07$ is usually accepted.
\cite{an07}, with new spectroscopic measurements, obtained a value of $[{\rm Fe}/{\rm H}] = 0.11 \pm 0.03$, also showing that the cluster is slightly metal 
rich. This fact has been overlooked in some studies that have used solar values for metallicity, and which therefore underestimate the cluster distance and 
overestimate its age.

The distance to Praesepe has been determined with many methods, yielding slight differences among the measured values. \cite{gatewood94} used the Multichannel
Astrometric Photometer (MAP) of the Thaw Refractor of the University of Pittsburgh to determine a weighted mean parallax of $\pi=5.21$ mas for five cluster member stars. 
The geometric method used by \cite{loktin00} determines the apparent variation of the angular diameter of the cluster as it moves along
the line of sight and estimates the distance to the cluster from it. The basic idea of this method is very similar to that of the convergent point method. 
The photometric distances (main-sequence fitting) seem to show a large scatter. We summarize the previous distance measurements and
the methods used to obtain the values in Table \ref{tab:mM}.
 
We determined the distance and age of the cluster by simultaneously fitting the distance modulus and the age with
isochrones. The photometry values we used were our best SDSS $g$ and $r$ band data, with the corrections explained
in \S 3. We did not include reddening in our color values, because it can be neglected towards Praesepe (E(B-V)=0.027 $\pm$ 0.004 mag; \cite{taylor06}).
Since the plotted CMD of Praesepe clearly showed a vertical trend at the later spectral type stars
at $g-r \approx 1.2$, we only fitted cluster member points with $g-r < 1.2$ (fitting the distance modulus to a
vertical trend is impossible and only adds errors to the fit). The isochrones for the fit were obtained from the Padova group 
website\footnote{http://pleiadi.pd.astro.it/}, where isochrones of any age and metallicity can be 
generated for a large number of photometric systems, such as the SDSS system \citep{girardi04}. These isochrones are similar to the empirical
isochrones produced by \cite{an07}.

Since the metallicity of the cluster is still debated, we fitted isochrone sets for all metallicities in the literature.
Assuming that metallicities are solar scaled, we set [Fe/H]=[M/H]. We fitted the following: 
[Fe/H]=0.13 \citep[Z=0.025;][]{boesgaard88}, [Fe/H]=0.11 \citep[Z=0.024;][]{an07} and [M/H]=0.2 \citep[Z=0.03;][]{an07}.
The two values from the \cite{an07} paper are from [Fe/H], determined from spectroscopy, and an [M/H]
value from isochrone fitting. We also fitted solar metallicity isochrones to show the errors they give in the 
age and distance determinations.

We calculated the best fit via Monte Carlo (i.e.\ bootstrap) method. We generated 10,000 new samples with the same number of sources as in the original
cluster member list. As with the bootstrap method, the members in the new samples were randomly picked from the original, resulting in
multiple picks of a few sources and null of others. The best fitting isochrones (as a function of age and distance) to these mock samples
were found by $\chi^2$ minimization. We computed $\chi^2$ from each fit as
\begin{equation}
\chi^2 = \sum(\Delta r^2 + \Delta (g-r)^2),
\end{equation}
where $\Delta r^2$ is the r magnitude difference while $\Delta(g-r)^2$ is the color difference from the closest point
of the isochrone model. By finding the closest point of the isochrones we not only fit the luminosity difference, but an actual distance from the 
isochrone, thus allowing points to be horizontally offset. We did not weight our fit by photometric errors, because the brightest members (that are most 
crucial in the age determination) did not have quoted errors, while the errors of the SDSS data cannot be trusted brighter than $14^{\rm th}$ magnitude. 
The means and errors in age and distance modulus of the best fitting isochrone (for all metalicities) were calculated from the distribution of solutions 
given by the bootstrap method. Following an initial fit, we removed stars that were further than 3$\sigma_{m-M}$ magnitude from the best fitting isochrones
and reran the Monte Carlo code. The best fit value is given as the arithmetic mean and its error as its standard deviation.

The 2D errors for the fits are shown in Figure \ref{fig:isofit}, both for the full and for the clipped samples. The histograms of the 
distance modulus and age fits are shown in Figure \ref{fig:histos}, both for the full and for the clipped samples also. The results of the fitting for the 
3$\sigma_{m-M}$ clipped sample are summarized in Table \ref{tab:distances} quoting the 1$\sigma$ errors. These errors are purely from the fitting procedure,
and do not include possible systematic errors such as those from isochrone models, reddening, extinction and photometry.

The best fitting isochrones for the four metallicities are shown in Figure \ref{fig:grr}. All isochrones seem to deviate from the observed 
trend at $g-r>1.2$ magnitude. This is either due to errors in the calculated isochrones or to the membership criteria of \cite{kraus07}, who used estimated 
T$_{\rm eff}$ and luminosity values from photometry fitted SEDs and theoretical Hertzsprung-Russell diagrams.

\begin{deluxetable}{llc}
\tablecolumns{3}
\tablewidth{0pt}
\tablecaption{The distance-modulus of Praesepe in the literature.\label{tab:mM}}
\tablehead{
\colhead{~~~~~~~~~~~~~~~~~~~~~~~~~~~~~~~~~~~~~~~~Reference~~~~~~~~~~~~~~~~~~~~~~~~~~~~~~~~~~~~~~~~} & \colhead{Method used} & \colhead{$m-M$} \\
& & \colhead{[mag]}}
\startdata
\cite{nissen88}\dotfill 	& Photometric 			& 6.05 \\
\cite{mermilliod90}\dotfill 	& Photometric 			& 6.2 \\
\cite{hauck81}\dotfill 		& Photometric\tablenotemark{a} 	& 6.26$\pm$0.23\\
\cite{vandenberg84}\dotfill 	& Photometric 			& 5.85 \\
\cite{an07}\dotfill 		& Photometric\tablenotemark{b} 	& 6.33 $\pm$ 0.04\\
\cite{gatewood94}\dotfill 	& Parallax 			& 6.42 $\pm$ 0.33\\
\cite{loktin00}\dotfill 	& Geometric 			& 6.16 $\pm$ 0.19\\
This paper\dotfill 		& Photometric 			& 6.267 $\pm$ 0.024
\enddata
\tablenotetext{a}{Using Lutz-Kelker corrections \citep{lutz73}.}
\tablenotetext{b}{Using empirically corrected isochrones.}
\end{deluxetable}

\begin{deluxetable}{llc}
\tablecolumns{3}
\tablewidth{0pt}
\tablecaption{The solutions for the fitting of isochrones via two parameters for solar and the metallicities found in the literature given with 1$\sigma$ errors.
These values are the ones determined after the 3$\sigma$ clipping iteration.\label{tab:distances}}
\tablehead{
\colhead{~~~~~~~~~~~~~~~~~~~~~~~~~~~~~~~~~~~~~~~~Metallicity~~~~~~~~~~~~~~~~~~~~~~~~~~~~~~~~~~~~~~~~} & \colhead{$m-M$} & \colhead{Age} \\
& \colhead{[mag]} & \colhead{[$\log t$]}}
\startdata
${\rm [Fe/H]} = 0.00$\dotfill                  & $6.012\pm0.020$ & $8.952\pm0.011$ \\
${\rm [Fe/H]} = 0.13$\tablenotemark{a}\dotfill & $6.153\pm0.022$ & $8.918\pm0.018$ \\
${\rm [Fe/H]} = 0.11$\tablenotemark{b}\dotfill & $6.179\pm0.022$ & $8.908\pm0.019$ \\
${\rm [M/H]}  = 0.20$\tablenotemark{b}\dotfill & $6.267\pm0.024$ & $8.879\pm0.020$ 
\enddata
\tablenotetext{a}{\cite{boesgaard88}}
\tablenotetext{b}{\cite{an07}}
\end{deluxetable}  
 
We adopted the metallicity of Z=0.03 \citep{an07} to give a final estimate of the cluster's age and distance. We chose this metallicity to ensure comparability, 
since \cite{an07} deduced it from isochrone fitting also. The distance modulus of our best fit for this metallicity is $m-M=6.267 \pm 0.024$ at 1$\sigma$ confidence, 
within errorbars of the value of \cite{an07} ($m-M=6.33 \pm 0.04$ mag). The errorbars on distance are small at 3$\sigma$ and comparable to the diameter of the 
cluster's central region ($\sim$ 6 pc). The age of the cluster is determined to be $\log t=8.879\pm0.020$ (757 $\pm$ 36 Myr) at 1$\sigma$ confidence. The errorbars 
on cluster age are significantly smaller than in papers before and help to pin down the decay trend at ages between 0.5 and 1 Gyr. The bootstrap Monte Carlo isochrone 
fitting method we introduce here turned out to be a very effective and successful way to determine cluster distance and age, and to estimate the errors of these parameters.

\clearpage
\LongTables
\begin{landscape} 
\setcounter{table}{0}
\begin{deluxetable}{lccrrrrrrrrc}
\tabletypesize{\scriptsize}
\tablecolumns{12}
\tablewidth{0pt}
\tablecaption{Photometry of Praesepe members in the [24] band \label{tab:solutions}}
\tablehead{
\colhead{\#} & 
\colhead{$\alpha_{2000}$} & 
\colhead{$\delta_{2000}$} & 
\colhead{g\tablenotemark{$\ast$}} &
\colhead{r\tablenotemark{$\ast$}} & 
\colhead{K$_{\rm 2MASS}$}& 
\colhead{[24]\tablenotemark{$\dagger$}} & 
\colhead{F$_{24}$\tablenotemark{$\dagger$}} & 
\colhead{$\mu_{\alpha}$} & 
\colhead{$\mu_{\delta}$} & 
\colhead{W\#\tablenotemark{$\ddagger$}} & 
\colhead{2MASS}\\
\colhead{} & 
\colhead{[h:m:s]} & 
\colhead{[$^{\circ}$:$'$:$''$]} & 
\colhead{[mag]} & 
\colhead{[mag]} & 
\colhead{[mag]} & 
\colhead{[mag]} & 
\colhead{[mJy]} & 
\colhead{[mas yr$^{-1}$]} &
\colhead{[mas yr$^{-1}$]} & 
& \colhead{}}
\startdata
1   & 8:36:29.83 & 18:57:56.52 & 9.57\phm{$^{\ast}$}  & 9.33$^{\ast}$	     &  8.30$\pm$0.01 & 8.23$\pm$0.05  &   3.54$\pm$ 0.15 & -34.60 & -12.60 & -   & 08362985+1857570\\
2   & 8:36:48.95 & 19:15:26.06 & 11.54$^{\ast}$       & 10.92$^{\ast}$	     &  9.69$\pm$0.02 & 9.35$\pm$0.06  &   1.26$\pm$ 0.07 & -36.30 & -12.80 & -   & 08364896+1915265\\
3   & 8:37:02.04 & 19:36:17.42 & 9.34\phm{$^{\ast}$}  & 9.06$^{\ast}$	     &  8.06$\pm$0.01 & 8.01$\pm$0.02  &   4.32$\pm$ 0.09 & -34.30 & -13.00 & -   & 08370203+1936171\\
4   & 8:37:16.35 & 19:29:11.58 & 14.67\phm{$^{\ast}$} & 13.38\phm{$^{\ast}$} & 10.47$\pm$0.02 & 10.06$\pm$0.07 &   0.65$\pm$ 0.04 & -34.70 & -15.40 & 267 & 08371635+1929103\\
5   & 8:37:18.29 & 19:41:56.33 & 11.75\phm{$^{\ast}$} & 11.23\phm{$^{\ast}$} &  9.80$\pm$0.02 & 9.69$\pm$0.05  &   0.93$\pm$ 0.04 & -37.20 & -15.20 & 268 & 08371829+1941564\\
6   & 8:37:26.51 & 19:29:13.06 & 14.48\phm{$^{\ast}$} & 13.33\phm{$^{\ast}$} & 10.83$\pm$0.02 & 10.77$\pm$0.15 &   0.34$\pm$ 0.05 & -42.70 & -14.40 & 274 & 08372638+1929128\\
7   & 8:37:27.58 & 19:37:03.29 & 11.97\phm{$^{\ast}$} & 11.40\phm{$^{\ast}$} &  9.81$\pm$0.02 & 9.75$\pm$0.06  &   0.87$\pm$ 0.04 & -34.10 & -12.60 & 277 & 08372755+1937033\\
8   & 8:37:27.95 & 19:33:45.25 & 9.91$^{\ast}$        & 9.64\phm{$^{\ast}$}  &  8.46$\pm$0.02 & 8.48$\pm$0.02  &   2.80$\pm$ 0.06 & -36.60 & -13.20 & 2   & 08372793+1933451\\
9   & 8:37:28.22 & 19:09:44.32 & 9.65\phm{$^{\ast}$}  & 9.37$^{\ast}$	     &  8.40$\pm$0.01 & 8.39$\pm$0.02  &   3.06$\pm$ 0.05 & -36.20 & -13.40 & 3   & 08372819+1909443\\
10  & 8:37:33.84 & 20:00:49.39 & 8.76$^{\ast}$        & 8.66$^{\ast}$	     &  7.95$\pm$0.03 & 7.95$\pm$0.02  &   4.57$\pm$ 0.10 & -35.70 & -13.10 & 5   & 08373381+2000492\\
11  & 8:37:36.33 & 19:15:53.96 & 14.05$^{\ast}$       & 13.05$^{\ast}$	     & 10.76$\pm$0.02 & 10.34$\pm$0.09 &   0.51$\pm$ 0.04 & -35.30 & -11.20 & 288 & 08373624+1915542\\
12  & 8:37:37.00 & 19:43:58.69 & 7.77$^{\ast}$        & 7.79$^{\ast}$	     &  7.29$\pm$0.01 & 7.30$\pm$0.01  &   8.31$\pm$ 0.09 & N/A    & N/A    & 6   & 08373699+1943585\\
13  & 8:37:40.71 & 19:31:06.38 & 8.29\phm{$^{\ast}$}  & 8.20$^{\ast}$	     &  7.66$\pm$0.01 & 7.63$\pm$0.01  &   6.14$\pm$ 0.08 & -34.80 & -12.50 & 10  & 08374070+1931063\\
14  & 8:37:42.36 & 19:08:01.57 & 10.05$^{\ast}$       & 9.75$^{\ast}$	     &  8.58$\pm$0.02 & 8.58$\pm$0.02  &   2.57$\pm$ 0.05 & -36.60 & -13.50 & 12  & 08374235+1908015\\
15  & 8:37:46.35 & 19:35:57.26 & 12.75\phm{$^{\ast}$} & 12.04\phm{$^{\ast}$} & 10.24$\pm$0.02 & 10.14$\pm$0.06 &   0.61$\pm$ 0.03 & -37.80 & -9.40  & 295 & 08374640+1935575\\
16  & 8:37:46.64 & 19:26:18.10 & 10.85$^{\ast}$       & 10.50\phm{$^{\ast}$} &  9.28$\pm$0.02 & 9.32$\pm$0.04  &   1.29$\pm$ 0.04 & -36.10 & -13.40 & 13  & 08374660+1926181\\
17  & 8:37:46.77 & 19:16:02.03 & 6.75$^{\ast}$        & 6.76$^{\ast}$	     &  6.17$\pm$0.01 & 6.12$\pm$0.01  &  24.79$\pm$ 0.26 & N/A    & N/A    & 14  & 08374675+1916020\\
18  & 8:37:47.30 & 19:06:24.01 & 12.71\phm{$^{\ast}$} & 11.96$^{\ast}$	     & 10.20$\pm$0.02 & 10.07$\pm$0.07 &   0.65$\pm$ 0.04 & -35.60 & -15.10 & 299 & 08374739+1906247\\
19  & 8:37:49.99 & 19:53:28.75 & 11.78$^{\ast}$       & 11.13\phm{$^{\ast}$} &  9.33$\pm$0.02 & 9.10$\pm$0.04  &   1.59$\pm$ 0.05 & -31.80 & -19.20 & 304 & 08374998+1953287\\
20  & 8:37:52.08 & 19:59:13.85 & 11.54\phm{$^{\ast}$} & 11.07\phm{$^{\ast}$} &  9.69$\pm$0.02 & 9.55$\pm$0.04  &   1.05$\pm$ 0.04 & -38.80 & -14.60 & 310 & 08375208+1959138\\
21  & 8:37:57.06 & 19:14:09.67 & 12.23\phm{$^{\ast}$} & 11.59\phm{$^{\ast}$} & 10.04$\pm$0.02 & 10.11$\pm$0.09 &   0.63$\pm$ 0.05 & -35.40 & -13.70 & 325 & 08375703+1914103\\
22  & 8:38:07.63 & 19:59:16.40 & 12.48\phm{$^{\ast}$} & 11.82\phm{$^{\ast}$} &  9.90$\pm$0.02 & 10.02$\pm$0.07 &   0.68$\pm$ 0.05 & -38.10 & -13.50 & 346 & 08380758+1959163\\
23  & 8:38:08.08 & 20:26:20.83 & 12.08$^{\ast}$       & 11.47$^{\ast}$	     &  9.93$\pm$0.02 & 10.03$\pm$0.17 &   0.68$\pm$ 0.11 & -36.40 & -14.40 & 347 & 08380808+2026223\\
24  & 8:38:14.11 & 19:47:23.82 & 15.56\phm{$^{\ast}$} & 14.17\phm{$^{\ast}$} & 10.91$\pm$0.04 & 10.00$\pm$0.13 &   0.69$\pm$ 0.08 & N/A    & N/A    & 358 & 08381421+1947234\\
25  & 8:38:14.28 & 19:21:55.37 & 11.20\phm{$^{\ast}$} & 10.31\phm{$^{\ast}$} &  9.19$\pm$0.02 & 9.12$\pm$0.03  &   1.56$\pm$ 0.05 & -35.00 & -13.70 & 21  & 08381427+1921552\\
26  & 8:38:23.16 & 20:12:26.60 & 8.01$^{\ast}$        & 7.71$^{\ast}$	     &  6.65$\pm$0.01 & 6.64$\pm$0.01  &  15.37$\pm$ 0.13 & N/A    & N/A    & -   & 08382311+2012263\\
27  & 8:38:24.31 & 20:06:21.92 & 10.80\phm{$^{\ast}$} & 10.40\phm{$^{\ast}$} &  9.18$\pm$0.02 & 9.23$\pm$0.03  &   1.41$\pm$ 0.04 & -36.30 & -13.10 & 24  & 08382429+2006217\\
28  & 8:38:29.70 & 19:51:45.83 & 14.67$^{\ast}$       & 13.53\phm{$^{\ast}$} & 10.93$\pm$0.02 & 10.56$\pm$0.14 &   0.41$\pm$ 0.06 & -40.10 & -13.20 & 27  & 08382963+1951450\\
29  & 8:38:32.18 & 19:27:55.04 & 10.46$^{\ast}$       & 9.65\phm{$^{\ast}$}  &  7.54$\pm$0.01 & 7.46$\pm$0.01  &   7.18$\pm$ 0.08 & N/A    & N/A    & -   & 08383216+1927548\\
30  & 8:38:34.27 & 19:51:36.90 & 9.68\phm{$^{\ast}$}  & 8.87\phm{$^{\ast}$}  &  6.69$\pm$0.01 & 6.66$\pm$0.01  &  15.03$\pm$ 0.18 & N/A    & N/A    & -   & 08383425+1951369\\
31  & 8:38:37.43 & 19:01:14.81 & 14.45\phm{$^{\ast}$} & 13.27\phm{$^{\ast}$} & 10.61$\pm$0.01 & 10.16$\pm$0.18 &   0.60$\pm$ 0.10 & -37.70 & -6.80  & 394 & 08383723+1901161\\
32  & 8:38:37.78 & 19:38:47.69 & 10.73$^{\ast}$       & 10.43\phm{$^{\ast}$} &  9.22$\pm$0.02 & 9.24$\pm$0.05  &   1.40$\pm$ 0.06 & N/A    & N/A    & -   & 08383776+1938480\\
33  & 8:38:37.88 & 19:59:23.14 & 8.16$^{\ast}$        & 8.16$^{\ast}$	     &  7.61$\pm$0.01 & 7.64$\pm$0.02  &   6.08$\pm$ 0.08 & -37.40 & -13.60 & 35  & 08383786+1959231\\
34  & 8:38:46.97 & 19:30:03.53 & 9.08\phm{$^{\ast}$}  & 8.95\phm{$^{\ast}$}  &  8.22$\pm$0.02 & 8.08$\pm$0.02  &   4.06$\pm$ 0.06 & -34.80 & -12.60 & 37  & 08384695+1930033\\
35  & 8:38:50.05 & 20:04:03.29 & 11.01\phm{$^{\ast}$} & 10.64\phm{$^{\ast}$} &  9.37$\pm$0.02 & 9.28$\pm$0.03  &   1.35$\pm$ 0.03 & -36.60 & -15.40 & 423 & 08385001+2004035\\
36  & 8:38:53.57 & 19:34:17.90 & 14.82$^{\ast}$       & 13.53$^{\ast}$	     & 10.96$\pm$0.02 & 10.92$\pm$0.17 &   0.30$\pm$ 0.05 & -40.90 & -21.30 & 432 & 08385354+1934170\\
37  & 8:38:55.07 & 19:11:54.02 & 10.84\phm{$^{\ast}$} & 9.87\phm{$^{\ast}$}  &  8.04$\pm$0.01 & 7.90$\pm$0.02  &   4.81$\pm$ 0.08 & N/A    & N/A    & -   & 08385506+1911539\\
38  & 8:39:01.89 & 20:00:19.62 & 12.48$^{\ast}$       & 11.45$^{\ast}$	     &  9.07$\pm$0.02 & 8.96$\pm$0.03  &   1.80$\pm$ 0.05 & N/A    & N/A    & -   & 08390185+2000194\\
39  & 8:39:02.27 & 19:19:35.36 & 12.83\phm{$^{\ast}$} & 12.06\phm{$^{\ast}$} & 10.26$\pm$0.02 & 10.20$\pm$0.08 &   0.57$\pm$ 0.04 & -36.60 & -10.50 & 448 & 08390228+1919343\\
40  & 8:39:02.84 & 19:43:28.99 & 9.48\phm{$^{\ast}$}  & 9.19$^{\ast}$	     &  8.12$\pm$0.01 & 8.08$\pm$0.02  &   4.06$\pm$ 0.06 & -35.80 & -11.20 & 46  & 08390283+1943289\\
41  & 8:39:03.24 & 20:02:35.12 & 15.18\phm{$^{\ast}$} & 13.86\phm{$^{\ast}$} & 11.05$\pm$0.02 & 10.88$\pm$0.13 &   0.31$\pm$ 0.04 & -40.70 & -14.30 & 47  & 08390321+2002376\\
42  & 8:39:03.60 & 19:59:59.24 & 8.33$^{\ast}$        & 8.32$^{\ast}$	     &  7.77$\pm$0.01 & 7.71$\pm$0.02  &   5.69$\pm$ 0.09 & -34.20 & -13.30 & 48  & 08390359+1959591\\
43  & 8:39:04.09 & 19:31:23.20 & 14.40\phm{$^{\ast}$} & 13.28\phm{$^{\ast}$} & 10.86$\pm$0.01 & 10.46$\pm$0.13 &   0.45$\pm$ 0.05 & -37.00 & -14.60 & 450 & 08390411+1931216\\
44  & 8:39:05.25 & 20:07:01.92 & 9.51\phm{$^{\ast}$}  & 9.31$^{\ast}$	     &  8.41$\pm$0.02 & 8.38$\pm$0.03  &   3.09$\pm$ 0.07 & -35.70 & -12.10 & 49  & 08390523+2007018\\
45  & 8:39:06.12 & 19:40:36.59 & 7.48$^{\ast}$        & 7.43$^{\ast}$	     &  6.71$\pm$0.01 & 6.73$\pm$0.01  &  14.03$\pm$ 0.17 & N/A    & N/A    & 50  & 08390612+1940364\\
46  & 8:39:06.55 & 19:00:36.68 & 13.83\phm{$^{\ast}$} & 13.09\phm{$^{\ast}$} & 11.30$\pm$0.02 & 11.16$\pm$0.30 &   0.24$\pm$ 0.07 & N/A    & N/A    & 457 & 08390649+1900360\\
47  & 8:39:09.11 & 19:35:32.68 & 8.54$^{\ast}$        & 8.49$^{\ast}$	     &  7.88$\pm$0.02 & 7.87$\pm$0.03  &   4.93$\pm$ 0.12 & -35.30 & -12.00 & 52  & 08390909+1935327\\
48  & 8:39:10.15 & 19:40:42.56 & 9.55$^{\ast}$        & 9.32$^{\ast}$	     &  8.41$\pm$0.01 & 8.40$\pm$0.02  &   3.03$\pm$ 0.04 & -36.10 & -13.70 & 54  & 08391014+1940423\\
49  & 8:39:12.20 & 19:06:56.45 & 10.86\phm{$^{\ast}$} & 10.41\phm{$^{\ast}$} &  9.26$\pm$0.02 & 9.18$\pm$0.05  &   1.47$\pm$ 0.07 & -37.00 & -13.40 & 57  & 08391217+1906561\\
50  & 8:39:15.05 & 20:12:39.35 & 11.61\phm{$^{\ast}$} & 11.13\phm{$^{\ast}$} &  9.65$\pm$0.02 & 9.63$\pm$0.06  &   0.97$\pm$ 0.05 & -35.20 & -14.70 & 477 & 08391499+2012388\\
51  & 8:39:19.77 & 20:03:10.91 & 9.78\phm{$^{\ast}$}  & 8.97\phm{$^{\ast}$}  &  7.08$\pm$0.01 & 7.02$\pm$0.01  &  10.80$\pm$ 0.14 & N/A    & N/A    & -   & 08391972+2003107\\
52  & 8:39:21.88 & 19:51:40.86 & 12.97$^{\ast}$       & 12.20\phm{$^{\ast}$} & 10.37$\pm$0.02 & 10.07$\pm$0.10 &   0.65$\pm$ 0.06 & -36.40 & -8.80  & 65  & 08392185+1951402\\
53  & 8:39:24.99 & 19:27:33.70 & 10.75\phm{$^{\ast}$} & 10.01\phm{$^{\ast}$} &  9.00$\pm$0.01 & 8.89$\pm$0.03  &   1.92$\pm$ 0.04 & -37.00 & -14.90 & 66  & 08392498+1927336\\
54  & 8:39:28.63 & 19:28:25.00 & 12.07$^{\ast}$       & 11.32$^{\ast}$	     &  9.53$\pm$0.02 & 9.40$\pm$0.05  &   1.21$\pm$ 0.06 & -36.10 & -10.80 & 506 & 08392858+1928251\\
55  & 8:39:29.42 & 19:47:11.51 & 13.09$^{\ast}$       & 12.29\phm{$^{\ast}$} & 10.06$\pm$0.01 & 9.94$\pm$0.22  &   0.73$\pm$ 0.14 & -38.90 & -9.00  & 69  & 08392940+1947118\\
56  & 8:39:30.44 & 20:04:08.69 & 10.68$^{\ast}$       & 10.11\phm{$^{\ast}$} &  8.81$\pm$0.01 & 8.74$\pm$0.02  &   2.22$\pm$ 0.03 & -35.80 & -13.40 & 70  & 08393042+2004087\\
57  & 8:39:33.44 & 20:10:10.52 & 9.79\phm{$^{\ast}$}  & 8.67\phm{$^{\ast}$}  &  6.13$\pm$0.01 & 6.00$\pm$0.02  &  27.56$\pm$ 0.38 & N/A    & N/A    & -   & 08393342+2010102\\
58  & 8:39:36.35 & 19:15:39.67 & 15.10$^{\ast}$       & 13.87\phm{$^{\ast}$} & 11.01$\pm$0.02 & 10.49$\pm$0.12 &   0.44$\pm$ 0.05 & -33.30 & -24.60 & 523 & 08393643+1915378\\
59  & 8:39:38.29 & 19:26:26.02 & 13.13\phm{$^{\ast}$} & 12.28\phm{$^{\ast}$} & 10.29$\pm$0.02 & 10.22$\pm$0.09 &   0.57$\pm$ 0.04 & -33.00 & -9.60  & 77  & 08393836+1926272\\
60  & 8:39:42.66 & 19:46:42.49 & 6.69$^{\ast}$        & 6.65$^{\ast}$	     &  6.00$\pm$0.01 & 5.98$\pm$0.01  &  28.02$\pm$ 0.37 & N/A    & N/A    & 79  & 08394265+1946425\\
61  & 8:39:42.81 & 20:05:10.46 & 7.75$^{\ast}$        & 7.73$^{\ast}$	     &  7.16$\pm$0.01 & 7.16$\pm$0.03  &   9.47$\pm$ 0.23 & N/A    & N/A    & 80  & 08394279+2005103\\
62  & 8:39:43.35 & 19:25:10.52 & 12.27\phm{$^{\ast}$} & 10.86\phm{$^{\ast}$} &  7.90$\pm$0.02 & 7.76$\pm$0.01  &   5.44$\pm$ 0.06 & N/A    & N/A    & -   & 08394333+1925121\\
63  & 8:39:44.68 & 19:16:30.94 & 7.68$^{\ast}$        & 7.69$^{\ast}$	     &  7.09$\pm$10.00 &7.15$\pm$0.01  &   9.53$\pm$ 0.10 & N/A    & N/A    & 82  & 08394466+1916308\\
64  & 8:39:45.78 & 19:22:01.06 & 10.93\phm{$^{\ast}$} & 10.50\phm{$^{\ast}$} &  9.26$\pm$0.02 & 9.39$\pm$0.04  &   1.21$\pm$ 0.05 & -35.40 & -12.80 & 83  & 08394575+1922011\\
65  & 8:39:50.74 & 19:32:26.92 & 7.06$^{\ast}$        & 6.26$^{\ast}$	     &  4.39$\pm$0.04 & 4.32$\pm$0.01  & 129.25$\pm$ 1.24 & N/A    & N/A    & 86  & 08395072+1932269\\
66  & 8:39:50.86 & 19:33:02.23 & 12.15$^{\ast}$       & 11.56$^{\ast}$	     & 10.00$\pm$0.02 & 9.77$\pm$0.06  &   0.86$\pm$ 0.05 & -36.10 & -13.90 & 87  & 08395084+1933020\\
67  & 8:39:52.35 & 19:18:45.61 & 10.68\phm{$^{\ast}$} & 10.07\phm{$^{\ast}$} &  9.01$\pm$0.02 & 8.90$\pm$0.03  &   1.91$\pm$ 0.05 & -34.80 & -14.30 & 89  & 08395234+1918455\\
68  & 8:39:55.08 & 20:03:54.47 & 10.37\phm{$^{\ast}$} & 10.02\phm{$^{\ast}$} &  8.96$\pm$0.02 & 8.86$\pm$0.04  &   1.99$\pm$ 0.07 & -37.50 & -13.90 & 93  & 08395506+2003541\\
69  & 8:39:56.51 & 19:33:10.91 & 7.32$^{\ast}$        & 7.33$^{\ast}$	     &  6.79$\pm$0.01 & 6.82$\pm$0.01  &  12.99$\pm$ 0.15 & N/A    & N/A    & 94  & 08395649+1933107\\
70  & 8:39:57.78 & 19:32:29.26 & 7.58$^{\ast}$        & 7.53$^{\ast}$	     &  7.01$\pm$0.02 & 7.01$\pm$0.01  &  10.84$\pm$ 0.09 & N/A    & N/A    & 96  & 08395777+1932293\\
71  & 8:39:58.09 & 19:12:05.98 & 9.71\phm{$^{\ast}$}  & 9.38\phm{$^{\ast}$}  &  8.48$\pm$0.02 & 8.36$\pm$0.02  &   3.15$\pm$ 0.07 & -37.40 & -12.50 & 97  & 08395807+1912058\\
72  & 8:39:58.40 & 20:09:29.99 & 8.71\phm{$^{\ast}$}  & 8.86\phm{$^{\ast}$}  &  8.10$\pm$0.01 & 8.03$\pm$0.07  &   4.26$\pm$ 0.28 & -36.00 & -13.80 & 98  & 08395838+2009298\\
73  & 8:39:59.10 & 20:01:53.15 & 9.35\phm{$^{\ast}$}  & 9.15\phm{$^{\ast}$}  &  8.21$\pm$0.02 & 8.22$\pm$0.01  &   3.56$\pm$ 0.05 & -36.40 & -16.20 & 99  & 08395908+2001532\\
74  & 8:39:59.19 & 19:40:08.58 & 9.86\phm{$^{\ast}$}  & 9.69\phm{$^{\ast}$}  &  8.78$\pm$0.02 & 8.73$\pm$0.03  &   2.24$\pm$ 0.06 & N/A    & N/A    & -   & 08395915+1940083\\
75  & 8:39:59.58 & 18:56:35.30 & 10.08$^{\ast}$       & 9.95\phm{$^{\ast}$}  &  9.30$\pm$0.01 & 9.32$\pm$0.05  &   1.29$\pm$ 0.06 & N/A    & N/A    & -   & 08395957+1856357\\
76  & 8:39:59.84 & number19:34:00.55 & 12.41$^{\ast}$       & 11.57$^{\ast}$	     &  9.48$\pm$0.02 & 9.31$\pm$0.04  &   1.31$\pm$ 0.05 & -33.80 & -12.20 & 565 & 08395983+1934003\\
77  & 8:40:00.01 & 19:34:39.86 & 13.35$^{\ast}$       & 12.51$^{\ast}$	     & 10.55$\pm$0.02 & 10.20$\pm$0.06 &   0.57$\pm$ 0.03 & -39.40 & -4.20  & 100 & 08395998+1934405\\
78  & 8:40:00.64 & 19:48:23.44 & 10.49$^{\ast}$       & 10.17$^{\ast}$	     &  9.08$\pm$0.02 & 8.97$\pm$0.03  &   1.80$\pm$ 0.05 & -36.30 & -13.10 & 101 & 08400062+1948235\\
79  & 8:40:01.32 & 20:08:08.38 & 9.82$^{\ast}$        & 9.57$^{\ast}$	     &  8.62$\pm$0.01 & 8.53$\pm$0.02  &   2.67$\pm$ 0.05 & -36.00 & -14.50 & 102 & 08400130+2008082\\
80  & 8:40:01.72 & 18:59:59.17 & 10.47\phm{$^{\ast}$} & 9.93\phm{$^{\ast}$}  &  8.70$\pm$0.02 & 8.48$\pm$0.03  &   2.81$\pm$ 0.07 & -36.50 & -11.70 & 103 & 08400171+1859595\\
81  & 8:40:04.20 & 19:47:04.24 & 12.11\phm{$^{\ast}$} & 11.54\phm{$^{\ast}$} & 10.00$\pm$0.02 & 10.05$\pm$0.10 &   0.66$\pm$ 0.06 & -33.00 & -13.70 & 576 & 08400416+1947039\\
82  & 8:40:04.92 & 19:43:45.48 & 9.95\phm{$^{\ast}$}  & 9.67\phm{$^{\ast}$}  &  8.65$\pm$0.02 & 8.54$\pm$0.03  &   2.65$\pm$ 0.06 & -36.10 & -12.50 & 106 & 08400491+1943452\\
83  & 8:40:05.70 & 19:01:30.18 & 13.20\phm{$^{\ast}$} & 12.31\phm{$^{\ast}$} & 10.01$\pm$0.02 & 9.90$\pm$0.07  &   0.76$\pm$ 0.05 & -35.70 & -12.20 & 578 & 08400571+1901307\\
84  & 8:40:06.28 & 19:27:14.80 & 10.55\phm{$^{\ast}$} & 10.10\phm{$^{\ast}$} &  8.87$\pm$0.02 & 8.80$\pm$0.03  &   2.10$\pm$ 0.05 & N/A    & N/A    & -   & 08400627+1927148\\
85  & 8:40:06.37 & 19:18:26.46 & 11.58$^{\ast}$       & 10.76\phm{$^{\ast}$} &  9.23$\pm$0.02 & 9.23$\pm$0.03  &   1.40$\pm$ 0.04 & -34.30 & -14.70 & 582 & 08400635+1918264\\
86  & 8:40:06.44 & 20:00:28.12 & 6.88$^{\ast}$        & 6.06$^{\ast}$	     &  4.20$\pm$0.02 & 4.13$\pm$0.01  & 153.97$\pm$ 1.35 & N/A    & N/A    & 111 & 08400643+2000280\\
87  & 8:40:09.74 & 19:37:17.83 & 12.54$^{\ast}$       & 11.71\phm{$^{\ast}$} & 10.13$\pm$0.02 & 10.04$\pm$0.08 &   0.67$\pm$ 0.05 & -33.90 & -10.50 & 114 & 08400968+1937170\\
88  & 8:40:11.46 & 19:58:16.21 & 6.78$^{\ast}$        & 6.71\phm{$^{\ast}$}  &  6.53$\pm$0.02 & 6.56$\pm$0.01  &  16.42$\pm$ 0.17 & N/A    & N/A    & 115 & 08401145+1958161\\
89  & 8:40:12.32 & 19:38:22.78 & 10.07$^{\ast}$       & 9.79$^{\ast}$	     &  8.67$\pm$0.02 & 8.56$\pm$0.03  &   2.62$\pm$ 0.07 & -36.90 & -14.50 & 116 & 08401231+1938222\\
90  & 8:40:13.45 & 19:46:45.08 & 13.75$^{\ast}$       & 12.79$^{\ast}$	     & 10.64$\pm$0.02 & 10.75$\pm$0.13 &   0.35$\pm$ 0.04 & -31.50 & -14.40 & 117 & 08401345+1946436\\
91  & 8:40:15.36 & 19:59:39.66 & 8.88$^{\ast}$        & 8.77$^{\ast}$	     &  8.04$\pm$9.99 & 8.03$\pm$0.02  &   4.27$\pm$ 0.06 & -35.80 & -12.30 & 119 & 08401535+1959394\\
92  & 8:40:15.59 & 19:27:29.84 & 14.61\phm{$^{\ast}$} & 13.47\phm{$^{\ast}$} & 10.69$\pm$0.02 & 10.52$\pm$0.12 &   0.43$\pm$ 0.05 & -36.30 & -8.50  & 601 & 08401549+1927310\\
93  & 8:40:15.72 & 19:54:54.07 & 13.15\phm{$^{\ast}$} & 12.29\phm{$^{\ast}$} & 10.01$\pm$0.02 & 9.92$\pm$0.06  &   0.75$\pm$ 0.04 & -38.00 & -13.20 & 120 & 08401571+1954542\\
94  & 8:40:17.63 & 19:47:15.14 & 10.20\phm{$^{\ast}$} & 9.73$^{\ast}$	     &  8.58$\pm$0.02 & 8.60$\pm$0.03  &   2.52$\pm$ 0.07 & -35.50 & -13.60 & 122 & 08401762+1947152\\
95  & 8:40:18.10 & 19:31:55.13 & 7.52$^{\ast}$        & 7.57$^{\ast}$	     &  7.16$\pm$0.01 & 7.18$\pm$0.01  &   9.34$\pm$ 0.08 & N/A    & N/A    & 123 & 08401810+1931552\\
96  & 8:40:18.97 & 20:11:31.16 & 13.59\phm{$^{\ast}$} & 12.38$^{\ast}$	     & 10.04$\pm$0.01 & 9.71$\pm$0.06  &   0.90$\pm$ 0.05 & -37.40 & -14.00 & 607 & 08401893+2011307\\
97  & 8:40:20.16 & 19:20:56.44 & 6.83$^{\ast}$        & 6.77$^{\ast}$	     &  6.04$\pm$0.01 & 6.01$\pm$0.01  &  27.23$\pm$ 0.31 & N/A    & N/A    & 125 & 08402013+1920564\\
98  & 8:40:20.75 & 19:41:12.23 & 7.68$^{\ast}$        & 7.69$^{\ast}$	     &  7.28$\pm$0.02 & 7.30$\pm$0.01  &   8.36$\pm$ 0.11 & N/A    & N/A    & 127 & 08402075+1941120\\
99  & 8:40:22.09 & 19:40:11.82 & 6.95$^{\ast}$        & 6.08$^{\ast}$	     &  4.18$\pm$0.03 & 4.07$\pm$0.01  & 162.87$\pm$ 1.71 & N/A    & N/A    & 128 & 08402209+1940116\\
100 & 8:40:22.33 & 20:06:24.88 & 10.26\phm{$^{\ast}$} & 9.97\phm{$^{\ast}$}  &  8.85$\pm$0.01 & 8.64$\pm$0.03  &   2.43$\pm$ 0.06 & -36.60 & -12.20 & 129 & 08402231+2006243\\
101 & 8:40:22.73 & 19:27:53.46 & 10.94$^{\ast}$       & 10.53$^{\ast}$	     &  9.34$\pm$0.02 & 9.25$\pm$0.04  &   1.39$\pm$ 0.05 & -37.80 & -13.30 & 131 & 08402271+1927531\\
102 & 8:40:23.29 & 19:40:23.95 & 10.61$^{\ast}$       & 10.20$^{\ast}$	     &  9.01$\pm$0.02 & 9.01$\pm$0.03  &   1.73$\pm$ 0.04 & -37.00 & -11.80 & 132 & 08402327+1940236\\
103 & 8:40:23.48 & 19:50:06.04 & 8.09$^{\ast}$        & 8.04\phm{$^{\ast}$}  &  7.59$\pm$0.01 & 7.55$\pm$0.02  &   6.65$\pm$ 0.10 & N/A    & N/A    & 133 & 08402347+1950059\\
104 & 8:40:25.55 & 19:28:32.92 & 9.75\phm{$^{\ast}$}  & 9.37\phm{$^{\ast}$}  &  8.76$\pm$0.02 & 8.71$\pm$0.03  &   2.28$\pm$ 0.06 & -36.80 & -13.30 & 134 & 08402554+1928328\\
105 & 8:40:26.14 & 19:41:11.33 & 9.50$^{\ast}$        & 9.27$^{\ast}$	     &  8.37$\pm$0.02 & 8.28$\pm$0.02  &   3.37$\pm$ 0.05 & -37.20 & -11.90 & 135 & 08402614+1941111\\
106 & 8:40:26.30 & 19:13:11.06 & 13.40\phm{$^{\ast}$} & 12.50\phm{$^{\ast}$} & 10.46$\pm$0.02 & 10.56$\pm$0.13 &   0.42$\pm$ 0.05 & -38.40 & -7.00  & 624 & 08402624+1913099\\
107 & 8:40:26.76 & 20:10:55.34 & 8.27$^{\ast}$        & 8.17$^{\ast}$	     &  7.43$\pm$0.01 & 7.38$\pm$0.01  &   7.73$\pm$ 0.09 & N/A    & N/A    & 136 & 08402675+2010552\\
108 & 8:40:27.03 & 19:32:41.42 & 6.27$^{\ast}$        & 6.31$^{\ast}$	     &  5.88$\pm$0.01 & 5.92$\pm$0.01  &  29.83$\pm$ 0.29 & N/A    & N/A    & 137 & 08402702+1932415\\
109 & 8:40:27.46 & 19:16:40.87 & 11.45\phm{$^{\ast}$} & 10.96$^{\ast}$	     &  9.65$\pm$0.02 & 9.58$\pm$0.04  &   1.02$\pm$ 0.04 & -33.30 & -12.10 & 628 & 08402743+1916409\\
110 & 8:40:27.52 & 19:39:20.05 & 13.77$^{\ast}$       & 12.83\phm{$^{\ast}$} & 10.69$\pm$0.02 & 10.84$\pm$0.12 &   0.32$\pm$ 0.04 & -33.40 & -12.20 & 138 & 08402751+1939197\\
111 & 8:40:28.68 & 20:18:44.86 & 12.04$^{\ast}$       & 11.35\phm{$^{\ast}$} &  9.46$\pm$0.02 & 9.42$\pm$0.07  &   1.18$\pm$ 0.07 & -37.40 & -15.90 & 631 & 08402863+2018449\\
112 & 8:40:31.72 & 19:51:01.84 & 11.98$^{\ast}$       & 11.38$^{\ast}$	     &  9.91$\pm$0.02 & 9.72$\pm$0.09  &   0.90$\pm$ 0.07 & -35.60 & -12.90 & 640 & 08403169+1951010\\
113 & 8:40:31.85 & 20:12:5.98  & 11.85$^{\ast}$       & 11.28\phm{$^{\ast}$} &  9.83$\pm$0.01 & 9.81$\pm$0.06  &   0.83$\pm$ 0.04 & -36.40 & -13.90 & 641 & 08403184+2012060\\
114 & 8:40:32.97 & 19:11:39.59 & 8.72\phm{$^{\ast}$}  & 8.55$^{\ast}$	     &  7.96$\pm$0.00 & 7.82$\pm$0.02  &   5.16$\pm$ 0.10 & -37.40 & -14.20 & 141 & 08403296+1911395\\
115 & 8:40:33.48 & 19:38:00.42 & 12.63$^{\ast}$       & 11.91$^{\ast}$	     & 10.17$\pm$0.02 & 10.05$\pm$0.08 &   0.66$\pm$ 0.05 & -38.90 & -10.60 & 142 & 08403347+1938009\\
116 & 8:40:39.25 & 19:13:41.88 & 7.82$^{\ast}$        & 7.80$^{\ast}$	     &  7.23$\pm$0.01 & 7.25$\pm$0.01  &   8.73$\pm$ 0.10 & N/A    & N/A    & 150 & 08403924+1913418\\
117 & 8:40:39.94 & 19:40:09.37 & 11.44$^{\ast}$       & 10.66\phm{$^{\ast}$} &  9.19$\pm$0.02 & 9.18$\pm$0.03  &   1.47$\pm$ 0.05 & -35.50 & -11.30 & 151 & 08403992+1940092\\
118 & 8:40:41.91 & 19:13:25.68 & 10.86$^{\ast}$       & 10.43$^{\ast}$	     &  9.06$\pm$0.02 & 9.04$\pm$0.03  &   1.68$\pm$ 0.04 & -35.90 & -13.00 & 153 & 08404189+1913255\\
119 & 8:40:42.51 & 19:33:57.85 & 11.66$^{\ast}$       & 11.11$^{\ast}$	     &  9.71$\pm$0.02 & 9.69$\pm$0.04  &   0.92$\pm$ 0.04 & -36.70 & -13.80 & 154 & 08404248+1933576\\
120 & 8:40:43.22 & 19:43:09.62 & 7.07$^{\ast}$        & 6.85\phm{$^{\ast}$}  &  6.33$\pm$0.01 & 6.33$\pm$0.01  &  20.45$\pm$ 0.14 & N/A    & N/A    & 156 & 08404321+1943095\\
121 & 8:40:46.09 & 19:18:34.67 & 9.79\phm{$^{\ast}$}  & 9.45\phm{$^{\ast}$}  &  8.53$\pm$0.02 & 8.49$\pm$0.02  &   2.79$\pm$ 0.05 & -37.10 & -13.20 & 158 & 08404608+1918346\\
122 & 8:40:47.23 & 19:32:37.64 & 10.87\phm{$^{\ast}$} & 10.08\phm{$^{\ast}$} &  8.20$\pm$0.01 & 8.10$\pm$0.01  &   3.99$\pm$ 0.05 & N/A    & N/A    & -   & 08404720+1932373\\
123 & 8:40:48.01 & 19:39:31.57 & 11.56\phm{$^{\ast}$} & 10.79\phm{$^{\ast}$} &  9.25$\pm$0.02 & 9.25$\pm$0.03  &   1.39$\pm$ 0.04 & -37.60 & -14.50 & 161 & 08404798+1939321\\
124 & 8:40:48.32 & 19:55:19.02 & 11.29\phm{$^{\ast}$} & 10.86\phm{$^{\ast}$} &  9.51$\pm$0.02 & 9.50$\pm$0.04  &   1.10$\pm$ 0.04 & -35.50 & -13.00 & 162 & 08404832+1955189\\
125 & 8:40:52.52 & 20:15:59.87 & 8.52\phm{$^{\ast}$}  & 8.47$^{\ast}$	     &  7.80$\pm$0.01 & 7.78$\pm$0.02  &   5.36$\pm$ 0.08 & -34.60 & -12.70 & 166 & 08405247+2015594\\
126 & 8:40:52.53 & 19:28:59.77 & 10.55\phm{$^{\ast}$} & 10.15\phm{$^{\ast}$} &  9.05$\pm$0.02 & 8.97$\pm$0.03  &   1.79$\pm$ 0.05 & -37.00 & -13.20 & 167 & 08405252+1928595\\
127 & 8:40:54.93 & 19:56:06.25 & 12.50$^{\ast}$       & 11.80$^{\ast}$	     & 10.13$\pm$0.02 & 10.16$\pm$0.09 &   0.60$\pm$ 0.05 & -37.20 & -14.90 & 677 & 08405487+1956067\\
128 & 8:40:56.29 & 19:34:49.26 & 6.76$^{\ast}$        & 6.79$^{\ast}$	     &  6.28$\pm$0.01 & 6.28$\pm$0.01  &  21.23$\pm$ 0.28 & N/A    & N/A    & 170 & 08405630+1934492\\
129 & 8:40:56.76 & 19:44:05.50 & 12.64$^{\ast}$       & 11.94$^{\ast}$	     & 10.21$\pm$0.02 & 10.13$\pm$0.11 &   0.61$\pm$ 0.06 & -36.10 & -11.10 & 171 & 08405669+1944052\\
130 & 8:40:56.95 & 19:56:05.57 & 8.79$^{\ast}$        & 8.67$^{\ast}$	     &  8.05$\pm$0.02 & 7.95$\pm$0.01  &   4.57$\pm$ 0.06 & -36.10 & -15.40 & 172 & 08405693+1956055\\
131 & 8:41:04.79 & 19:31:22.94 & 11.35\phm{$^{\ast}$} & 10.60\phm{$^{\ast}$} &  8.75$\pm$0.02 & 8.68$\pm$0.03  &   2.33$\pm$ 0.06 & N/A    & N/A    & -   & 08410478+1931225\\
132 & 8:41:07.34 & 19:26:48.08 & 13.01\phm{$^{\ast}$} & 12.12\phm{$^{\ast}$} & 10.29$\pm$0.02 & 10.30$\pm$0.20 &   0.53$\pm$ 0.09 & -43.70 & -8.10  & 176 & 08410725+1926489\\
133 & 8:41:07.39 & 19:04:16.43 & 10.56\phm{$^{\ast}$} & 10.04\phm{$^{\ast}$} &  8.64$\pm$0.02 & 8.53$\pm$0.03  &   2.68$\pm$ 0.06 & -39.90 & -14.10 & 177 & 08410737+1904164\\
134 & 8:41:09.61 & 19:51:18.32 & 11.01$^{\ast}$       & 10.50$^{\ast}$	     &  8.94$\pm$0.02 & 8.61$\pm$0.03  &   2.48$\pm$ 0.06 & -36.70 & -13.90 & 179 & 08410961+1951186\\
135 & 8:41:09.82 & 19:56:07.04 & 14.26$^{\ast}$       & 13.19\phm{$^{\ast}$} & 10.76$\pm$0.02 & 10.67$\pm$0.10 &   0.37$\pm$ 0.04 & -42.40 & -15.10 & 180 & 08410979+1956072\\
136 & 8:41:10.02 & 19:30:32.18 & 10.35$^{\ast}$       & 9.98\phm{$^{\ast}$}  &  8.91$\pm$0.02 & 8.83$\pm$0.13  &   2.04$\pm$ 0.23 & -36.90 & -12.00 & 182 & 08411002+1930322\\
137 & 8:41:10.32 & 19:49:07.10 & 11.84$^{\ast}$       & 11.29$^{\ast}$	     &  9.75$\pm$0.02 & 9.62$\pm$0.05  &   0.98$\pm$ 0.04 & -36.50 & -13.20 & 183 & 08411031+1949071\\
138 & 8:41:10.70 & 19:49:46.38 & 9.06\phm{$^{\ast}$}  & 8.86\phm{$^{\ast}$}  &  8.19$\pm$0.01 & 8.19$\pm$0.02  &   3.66$\pm$ 0.06 & -37.80 & -13.70 & 184 & 08411067+1949465\\
139 & 8:41:13.04 & 19:32:34.26 & 15.06\phm{$^{\ast}$} & 13.73\phm{$^{\ast}$} & 10.35$\pm$0.02 & 10.22$\pm$0.08 &   0.56$\pm$ 0.05 & -37.60 & -9.70  & 709 & 08411319+1932349\\
140 & 8:41:13.80 & 19:55:19.24 & 8.35$^{\ast}$        & 8.35$^{\ast}$	     &  7.77$\pm$0.01 & 7.69$\pm$0.02  &   5.82$\pm$ 0.09 & -36.90 & -12.60 & 188 & 08411377+1955191\\
141 & 8:41:15.43 & 20:02:15.04 & 14.99$^{\ast}$       & 13.78$^{\ast}$	     & 11.02$\pm$0.02 & 10.79$\pm$0.11 &   0.34$\pm$ 0.03 & -37.30 & -11.90 & 189 & 08411541+2002160\\
142 & 8:41:16.04 & 19:44:54.13 & 14.06\phm{$^{\ast}$} & 13.44\phm{$^{\ast}$} & 11.71$\pm$0.02 & 10.50$\pm$0.09 &   0.44$\pm$ 0.03 & N/A    & N/A    & 190 & 08411602+1944514\\
143 & 8:41:18.42 & 19:15:39.38 & 7.95$^{\ast}$        & 7.90$^{\ast}$	     &  7.29$\pm$0.02 & 7.04$\pm$0.01  &  10.61$\pm$ 0.12 & -37.40 & -12.90 & 192 & 08411840+1915394\\
144 & 8:41:19.96 & 19:38:04.20 & 14.20\phm{$^{\ast}$} & 13.09\phm{$^{\ast}$} & 10.76$\pm$0.02 & 10.22$\pm$0.17 &   0.56$\pm$ 0.09 & -36.30 & -12.00 & 721 & 08411992+1938047\\
145 & 8:41:22.48 & 18:56:00.17 & 13.49\phm{$^{\ast}$} & 12.60\phm{$^{\ast}$} & 10.54$\pm$0.02 & 10.20$\pm$0.15 &   0.58$\pm$ 0.08 & -34.00 & -9.90  & 726 & 08412258+1856020\\
146 & 8:41:23.93 & 20:14:57.30 & 15.46$^{\ast}$       & 14.11$^{\ast}$	     & 10.78$\pm$0.02 & 10.11$\pm$0.13 &   0.63$\pm$ 0.08 & N/A    & N/A    & 194 & 08412390+2014572\\
147 & 8:41:25.89 & 19:56:36.85 & 10.92$^{\ast}$       & 10.55\phm{$^{\ast}$} &  9.33$\pm$0.02 & 9.37$\pm$0.04  &   1.24$\pm$ 0.05 & -36.30 & -13.70 & 195 & 08412584+1956369\\
148 & 8:41:26.98 & 19:32:32.71 & 10.05\phm{$^{\ast}$} & 9.73$^{\ast}$	     &  8.72$\pm$0.02 & 8.79$\pm$0.03  &   2.12$\pm$ 0.06 & -37.30 & -12.40 & 196 & 08412698+1932329\\
149 & 8:41:28.65 & 19:44:49.13 & 11.43\phm{$^{\ast}$} & 10.76\phm{$^{\ast}$} &  9.47$\pm$0.02 & 9.44$\pm$0.13  &   1.16$\pm$ 0.14 & -39.00 & -13.50 & 198 & 08412869+1944481\\
150 & 8:41:33.89 & 19:58:08.83 & 12.07\phm{$^{\ast}$} & 11.47$^{\ast}$	     &  9.93$\pm$0.01 & 9.94$\pm$0.07  &   0.73$\pm$ 0.05 & -39.40 & -14.40 & 201 & 08413384+1958087\\
151 & 8:41:35.09 & 19:39:45.04 & 9.17$^{\ast}$        & 8.12\phm{$^{\ast}$}  &  5.96$\pm$0.02 & 5.86$\pm$0.01  &  31.26$\pm$ 0.30 & N/A    & N/A    & -   & 08413506+1939449\\
152 & 8:41:35.90 & 19:06:25.16 & 14.84\phm{$^{\ast}$} & 13.59\phm{$^{\ast}$} & 10.98$\pm$0.02 & 10.55$\pm$0.11 &   0.42$\pm$ 0.04 & -31.10 & -9.60  & 751 & 08413599+1906255\\
153 & 8:41:36.20 & 19:08:33.58 & 9.57\phm{$^{\ast}$}  & 9.23\phm{$^{\ast}$}  &  8.35$\pm$0.02 & 8.34$\pm$0.02  &   3.21$\pm$ 0.05 & -36.00 & -14.30 & 204 & 08413620+1908335\\
154 & 8:41:37.43 & 19:31:13.08 & 14.19\phm{$^{\ast}$} & 13.09\phm{$^{\ast}$} & 10.73$\pm$0.01 & 10.50$\pm$0.14 &   0.44$\pm$ 0.06 & -40.90 & -12.00 & 758 & 08413741+1931140\\
155 & 8:41:42.31 & 19:39:37.98 & 9.72$^{\ast}$        & 9.50$^{\ast}$	     &  8.48$\pm$0.02 & 8.38$\pm$0.03  &   3.09$\pm$ 0.08 & -37.30 & -13.80 & 206 & 08414229+1939379\\
156 & 8:41:43.68 & 19:57:43.85 & 12.73\phm{$^{\ast}$} & 12.05\phm{$^{\ast}$} & 10.26$\pm$0.02 & 10.15$\pm$0.10 &   0.60$\pm$ 0.10 & -40.50 & -13.10 & 769 & 08414368+1957437\\
157 & 8:41:43.85 & 20:13:37.06 & 10.69\phm{$^{\ast}$} & 10.34\phm{$^{\ast}$} &  9.14$\pm$0.01 & 8.99$\pm$0.04  &   1.76$\pm$ 0.07 & -37.40 & -15.70 & 771 & 08414382+2013368\\
158 & 8:41:45.49 & 19:16:02.17 & 10.35\phm{$^{\ast}$} & 9.98$^{\ast}$	     &  8.93$\pm$0.02 & 8.92$\pm$0.03  &   1.88$\pm$ 0.05 & -38.10 & -13.20 & 208 & 08414549+1916023\\
159 & 8:41:47.74 & 19:24:43.88 & 11.66$^{\ast}$       & 11.28\phm{$^{\ast}$} & 10.10$\pm$0.02 & 9.96$\pm$0.07  &   0.72$\pm$ 0.05 & -30.30 & -9.50  & -   & 08414776+1924439\\
160 & 8:41:48.24 & 19:27:30.49 & 14.28$^{\ast}$       & 13.24\phm{$^{\ast}$} & 10.73$\pm$0.01 & 10.62$\pm$0.10 &   0.39$\pm$ 0.04 & -40.80 & -9.80  & 774 & 08414818+1927312\\
161 & 8:41:49.34 & 19:11:47.51 & 15.24\phm{$^{\ast}$} & 13.90\phm{$^{\ast}$} & 10.83$\pm$0.01 & 10.55$\pm$0.11 &   0.42$\pm$ 0.04 & -33.90 & -10.80 & 776 & 08414934+1911471\\
162 & 8:41:50.09 & 19:52:27.19 & 7.37$^{\ast}$        & 6.56$^{\ast}$	     &  4.68$\pm$0.00 & 4.63$\pm$0.01  &  97.51$\pm$ 1.02 & N/A    & N/A    & 212 & 08415008+1952270\\
163 & 8:41:51.98 & 20:10:01.99 & 12.44$^{\ast}$       & 11.76$^{\ast}$	     & 10.09$\pm$0.02 & 9.78$\pm$0.08  &   0.85$\pm$ 0.07 & -40.60 & -15.70 & 213 & 08415199+2010013\\
164 & 8:41:53.16 & 20:09:34.16 & 8.61$^{\ast}$        & 8.51$^{\ast}$	     &  7.79$\pm$0.01 & 7.71$\pm$0.01  &   5.71$\pm$ 0.07 & -38.20 & -13.70 & 214 & 08415314+2009340\\
165 & 8:41:54.37 & 19:15:27.14 & 11.65\phm{$^{\ast}$} & 11.03\phm{$^{\ast}$} &  9.64$\pm$0.02 & 9.56$\pm$0.05  &   1.04$\pm$ 0.05 & -34.80 & -13.20 & 215 & 08415437+1915266\\
166 & 8:41:55.90 & 19:41:22.96 & 11.39\phm{$^{\ast}$} & 10.86\phm{$^{\ast}$} &  9.54$\pm$0.01 & 9.59$\pm$0.07  &   1.02$\pm$ 0.07 & -37.60 & -12.10 & 217 & 08415587+1941229\\
167 & 8:41:57.84 & 18:54:42.08 & 9.66$^{\ast}$        & 9.35\phm{$^{\ast}$}  &  8.43$\pm$0.02 & 8.43$\pm$0.03  &   2.94$\pm$ 0.07 & -34.30 & -11.10 & 218 & 08415782+1854422\\
168 & 8:41:58.86 & 20:06:26.82 & 13.77$^{\ast}$       & 12.83\phm{$^{\ast}$} & 10.60$\pm$0.01 & 10.32$\pm$0.16 &   0.52$\pm$ 0.08 & -40.30 & -11.80 & 792 & 08415884+2006272\\
169 & 8:42:05.50 & 19:35:57.95 & 11.07\phm{$^{\ast}$} & 10.33$^{\ast}$	     &  8.38$\pm$0.03 & 8.30$\pm$0.02  &   3.32$\pm$ 0.05 & N/A    & N/A    & -   & 08420547+1935585\\
170 & 8:42:06.51 & 19:24:40.72 & 7.96$^{\ast}$        & 7.97$^{\ast}$	     &  7.43$\pm$0.02 & 7.39$\pm$0.01  &   7.64$\pm$ 0.06 & -38.40 & -12.10 & 223 & 08420650+1924405\\
171 & 8:42:10.79 & 18:56:03.62 & 7.92$^{\ast}$        & 7.93$^{\ast}$	     &  7.35$\pm$0.01 & 7.35$\pm$0.02  &   7.95$\pm$ 0.11 & -34.10 & -12.10 & 224 & 08421080+1856037\\
172 & 8:42:11.50 & 19:16:36.37 & 12.57$^{\ast}$       & 11.85$^{\ast}$	     & 10.17$\pm$0.01 & 9.99$\pm$0.06  &   0.70$\pm$ 0.04 & -37.60 & -10.00 & 817 & 08421149+1916373\\
173 & 8:42:12.34 & 19:12:48.20 & 14.32\phm{$^{\ast}$} & 13.20\phm{$^{\ast}$} & 10.83$\pm$0.01 & 10.56$\pm$0.10 &   0.41$\pm$ 0.04 & -33.50 & -9.60  & 822 & 08421233+1912488\\
174 & 8:42:12.85 & 19:16:03.79 & 14.10$^{\ast}$       & 13.07\phm{$^{\ast}$} & 10.48$\pm$0.01 & 10.11$\pm$0.06 &   0.63$\pm$ 0.04 & -33.00 & -11.70 & 824 & 08421285+1916040\\
175 & 8:42:15.50 & 19:41:15.47 & 10.12\phm{$^{\ast}$} & 9.78\phm{$^{\ast}$}  &  8.77$\pm$0.01 & 8.72$\pm$0.02  &   2.24$\pm$ 0.05 & -37.60 & -15.00 & 226 & 08421549+1941156\\
176 & 8:42:18.85 & 20:24:36.22 & 12.59\phm{$^{\ast}$} & 11.91\phm{$^{\ast}$} & 10.19$\pm$0.02 & 10.05$\pm$0.12 &   0.66$\pm$ 0.08 & N/A    & N/A    & 839 & 08421883+2024350\\
177 & 8:42:20.16 & 20:02:11.72 & 9.91\phm{$^{\ast}$}  & 9.56\phm{$^{\ast}$}  &  8.41$\pm$0.02 & 8.40$\pm$0.02  &   3.03$\pm$ 0.06 & -35.70 & -15.60 & 228 & 08422012+2002117\\
178 & 8:42:21.62 & 20:10:53.72 & 9.32\phm{$^{\ast}$}  & 9.13\phm{$^{\ast}$}  &  8.28$\pm$0.01 & 8.13$\pm$0.02  &   3.89$\pm$ 0.06 & -36.80 & -14.40 & 229 & 08422162+2010539\\
179 & 8:42:24.74 & 19:35:17.27 & 11.21\phm{$^{\ast}$} & 10.83$^{\ast}$	     &  9.48$\pm$0.01 & 9.34$\pm$0.04  &   1.27$\pm$ 0.05 & N/A    & N/A    & -   & 08422471+1935175\\
180 & 8:42:32.27 & 19:23:46.25 & 11.38\phm{$^{\ast}$} & 10.84\phm{$^{\ast}$} &  9.46$\pm$0.01 & 9.50$\pm$0.06  &   1.10$\pm$ 0.06 & -36.50 & -12.50 & 232 & 08423225+1923463\\
181 & 8:42:40.19 & 19:07:58.87 & 12.55$^{\ast}$       & 11.86$^{\ast}$	     & 10.19$\pm$0.01 & 9.83$\pm$0.06  &   0.81$\pm$ 0.04 & -35.30 & -10.90 & 863 & 08424021+1907590\\
182 & 8:42:40.73 & 19:32:35.34 & 10.03\phm{$^{\ast}$} & 9.66\phm{$^{\ast}$}  &  8.72$\pm$0.02 & 8.67$\pm$0.03  &   2.36$\pm$ 0.07 & -38.40 & -12.70 & 235 & 08424071+1932354\\
183 & 8:42:42.51 & 19:05:59.78 & 12.04$^{\ast}$       & 11.38\phm{$^{\ast}$} &  9.88$\pm$0.02 & 9.88$\pm$0.06  &   0.78$\pm$ 0.04 & -37.40 & -13.50 & 236 & 08424250+1905589\\
184 & 8:42:43.72 & 19:37:23.52 & 12.76\phm{$^{\ast}$} & 11.76\phm{$^{\ast}$} &  9.80$\pm$0.02 & 9.64$\pm$0.05  &   0.97$\pm$ 0.04 & -36.40 & -14.20 & 868 & 08424372+1937234\\
185 & 8:42:44.44 & 19:34:48.11 & 10.09\phm{$^{\ast}$} & 9.53\phm{$^{\ast}$}  &  8.63$\pm$0.02 & 8.48$\pm$0.02  &   2.82$\pm$ 0.05 & -38.20 & -13.50 & 238 & 08424441+1934479\\
186 & 8:43:00.59 & 20:20:15.79 & 11.76\phm{$^{\ast}$} & 11.24\phm{$^{\ast}$} &  9.77$\pm$0.02 & 9.57$\pm$0.06  &   1.03$\pm$ 0.06 & -37.30 & -16.00 & 887 & 08430055+2020161\\
187 & 8:43:05.96 & 19:26:15.36 & 10.25\phm{$^{\ast}$} & 9.65\phm{$^{\ast}$}  &  8.46$\pm$0.02 & 8.40$\pm$0.01  &   3.02$\pm$ 0.04 & -36.60 & -13.80 & 248 & 08430593+1926152\\
188 & 8:43:08.24 & 19:42:47.59 & 13.92\phm{$^{\ast}$} & 12.89\phm{$^{\ast}$} & 10.67$\pm$0.01 & 10.15$\pm$0.09 &   0.60$\pm$ 0.05 & -33.70 & -11.60 & 899 & 08430822+1942475\\
189 & 8:43:10.82 & 19:31:33.64 & 12.20\phm{$^{\ast}$} & 11.58\phm{$^{\ast}$} & 10.01$\pm$0.02 & 10.49$\pm$0.22 &   0.44$\pm$ 0.09 & -38.50 & -17.50 & 902 & 08431076+1931346\\
190 & 8:43:20.20 & 19:46:08.58 & 11.05\phm{$^{\ast}$} & 10.62\phm{$^{\ast}$} &  9.36$\pm$0.02 & 9.26$\pm$0.04  &   1.37$\pm$ 0.05 & -39.40 & -13.40 & 255 & 08432019+1946086\\
191 & 8:43:32.42 & 19:44:38.00 & 12.92\phm{$^{\ast}$} & 12.01\phm{$^{\ast}$} & 10.22$\pm$0.02 & 9.94$\pm$0.08  &   0.73$\pm$ 0.05 & -40.10 & -16.50 & 919 & 08433239+1944378\\
192 & 8:43:35.56 & 20:11:22.63 & 10.29\phm{$^{\ast}$} & 9.99\phm{$^{\ast}$}  &  8.92$\pm$0.02 & 8.95$\pm$0.03  &   1.82$\pm$ 0.05 & -39.30 & -14.70 & 257 & 08433553+2011225\\
193 & 8:44:07.37 & 20:04:36.23 & 10.30\phm{$^{\ast}$} & 10.05\phm{$^{\ast}$} &  9.06$\pm$0.01 & 8.95$\pm$0.03  &   1.82$\pm$ 0.04 & N/A    & N/A    & -   & 08440734+2004369
\enddata
\tablenotetext{$\ast$}{The r and/or g magnitudes marked with a star were calculated from B and V magnitudes as described in \S 3, while the rest are the original 
SDSS values.}
\tablenotetext{$\dagger$}{The [24] magnitudes are the ones that were calibrated to the 2MASS K$_{S}$ magnitudes, while the mJy values in the F$_{24}$ column are
the original flux values.}
\tablenotetext{$\ddagger$}{The numbers in this column represent the numbering of \cite{wang95}.}
\end{deluxetable}
\clearpage
\end{landscape}

\end{document}